\documentclass[a4paper]{ustthesis}

\title{Novel Blockchain-based Protocols for Electronic Voting and Auctions}  % Title of the thesis.
\author{Zhaorun Lin}     % Author of the thesis.
\degree{\MPhil}             % Degree for which the thesis is. Options: \AM \MSc \MPhil \PhD
\stage{\Thesis}              % Stage of PhD document; use \Thesis for all other degree. Options: \PQE \Proposal \Thesis
\subject{Computer Science and Engineering} % Subject of the Degree.
\department{Department of Computer Science and Engineering}       % Department to which the thesis is submitted.
\advisor{Prof. Amir Kafshdar Goharshady}     % Supervisor. Additional co-supervisor can be added using \member
\coadvisor{Prof. Jiasi Shen}
\depthead{Prof. Xiaofang Zhou}     % department head.
\defencedate{2025}{08}{18}     % \defencedate{year}{month}{day}.

\usepackage{algorithm}
\usepackage{algorithmicx}
\usepackage{makecell}
\usepackage{longtable}
\usepackage{url}
\usepackage{multirow}
\usepackage{graphics}
\usepackage{graphicx}
\usepackage{amsmath}
\usepackage{cases}
\usepackage{amssymb,amsfonts,bm}
\usepackage{colortbl}
\usepackage{xcolor}
\usepackage{times}
\usepackage{array}
\usepackage{hyperref}
\usepackage{pdflscape}
\usepackage{bbding}
\usepackage{tikz}
\usepackage{pgfplots}
\usepackage{textcomp}
\usepackage{float}
\usepackage{paralist}
\usepackage[T1]{fontenc}
\usepackage[square,numbers]{natbib}
\usepackage[symbol]{footmisc}
\usepackage[noend]{algpseudocode}
\usepackage[normalem]{ulem}

\DeclareMathOperator*{\argmax}{argmax}      % for argmax

\newcommand{\pk}{\textsf{pk}}
\newcommand{\sk}{\textsf{sk}}
\newcommand{\compatible}{\rightleftharpoons}
\newcommand{\func}[1]{\textsc{#1}}

\newcolumntype{C}[1]{>{\centering\arraybackslash}m{#1}}
\renewcommand{\paragraph}[1]{\smallskip\noindent\textbf{\emph{#1.}}}

\graphicspath{ {./figure/} }

\begin{document}
%\begin{CJK}{UTF8}{song}  % Bitstream Cyber Bit song ti

%\begin{CJK*}{UTF8}{gbsn} % Arphic song ti

%%%%%%%%%%%%%%%%%%%%%%%%%%%%%%%%%%%%%%%%%%%%%%%%%%%%%%%%%%%%%%%%%%%%%%%%%
%                                                                       %
% Now the actual Thesis. The order of output MUST be followed:          %
%                                                                       %
%    1) TITLEPAGE                                                       %
%                                                                       %
% The \maketitle command generates the Title page as well as the        %
% Signature page.                                                       %
%                                                                       %
%%%%%%%%%%%%%%%%%%%%%%%%%%%%%%%%%%%%%%%%%%%%%%%%%%%%%%%%%%%%%%%%%%%%%%%%%

\maketitle

%%%%%%%%%%%%%%%%%%%%%%%%%%%%%%%%%%%%%%%%%%%%%%%%%%%%%%%%%%%%%%%%%%%%%%%%%
%                                                                       %
%     2) DEDICATION (Optional)                                          %
%                                                                       %
% The \dedication and \enddedication commands are optional. If          %
% specified it generates a page for dedication.                         %
%
%%%%%%%%%%%%%%%%%%%%%%%%%%%%%%%%%%%%%%%%%%%%%%%%%%%%%%%%%%%%%%%%%%%%%%%%%

\dedication
% This is an optional section.
\begin{center}
    To \textit{Qiquan Lin, Libing Lu, and Ziqi Lin}, my dearest family.
\end{center}
\enddedication
\newpage

%%%%%%%%%%%%%%%%%%%%%%%%%%%%%%%%%%%%%%%%%%%%%%%%%%%%%%%%%%%%%%%%%%%%%%%%%
%                                                                       %
%     3) ACKNOWLEDGMENTS                                                %
%                                                                       %
% \acknowledgments and \endacknowledgments defines the                  %
% Acknowledgments of the author of the Thesis.                          %
%                                                                       %
%%%%%%%%%%%%%%%%%%%%%%%%%%%%%%%%%%%%%%%%%%%%%%%%%%%%%%%%%%%%%%%%%%%%%%%%%

\acknowledgments
First and foremost, I want to express my utmost gratitude towards my supervisor, Prof. Amir Kafshdar Goharshady, whom I am extremely fortunate to have been advised by him. In the last semester of my Bachelor's degree, I was eagerly seeking opportunities in both academia and industry. Amir, who saw my talent, offered me a unique path to start my research in ALPACAS. Accepting this offer helped reshape my life. During my MPhil, Amir provided unselfish guidance, not only in my academic but also professional life. He helped enormously in my research, career choices, and search for PhDs and jobs. Similarly, I want to extend my warmest thanks to my co-supervisor, Jiasi Shen, who generously agreed to co-supervise and support me after Amir's move to the University of Oxford.

I would also like to thank the gifted researchers, who are also my friends, at the ALPACAS Research Group. A special thanks to Jonas Ballweg, Togzhan Barakbayeva, Soroush Farokhnia, and Pingjiang Li, who I had the chance to brainstorm with them on several projects, and produce two academic papers. I also want to thank Xuran Cai, Zhuo Cai, Giovanna Kobus Conrado, Singh Hitarth, Pavel Hudec, Kerim Kochekov, Chun Kit Lam, Sergei Novozhilov, Tian Shu, and Ahmed Zaher for their accompaniment in our inspirational reading group and gatherings.

I could not have had a fruitful MPhil life without my friends. I want to thank Bingxuan Li and Zhuohao Yin, my dearest roommates, for every game night when I enjoyed sitting through their cries and roars after they lost against me. I am also thankful to Runzhao Xu for every basketball game we have played together and for being there in my most difficult times. As I always said, my friends are what balanced my research and life.

I want to thank my parents, Qiquan Lin and Libing Lu for bringing me into this world, raising me up, funding my education, giving me every valuable life lesson, and so much more. I would also like to thank my younger sister, Ziqi Lin, for trying to advise me in my academic and professional life even though she is utterly uncomprehending about what I am doing. I could not have made it this far without them.

Finally, I want to thank Prof. Siu-Wing Cheng and Prof. Raymond Chi Wing Wong for kindly agreeing to join my Thesis Examination Committee and to review my thesis.

Following the norms of the ALPACAS Research Group, the names in this acknowledgment are of course listed in alphabetical order.
\endacknowledgments
\newpage

%%%%%%%%%%%%%%%%%%%%%%%%%%%%%%%%%%%%%%%%%%%%%%%%%%%%%%%%%%%%%%%%%%%%%%%%%
%                                                                       %
%     4) TABLE OF CONTENTS                                              %
%                                                                       %
%%%%%%%%%%%%%%%%%%%%%%%%%%%%%%%%%%%%%%%%%%%%%%%%%%%%%%%%%%%%%%%%%%%%%%%%%

\tableofcontents

%%%%%%%%%%%%%%%%%%%%%%%%%%%%%%%%%%%%%%%%%%%%%%%%%%%%%%%%%%%%%%%%%%%%%%%%%
%                                                                       %
%     5) LIST OF FIGURES (If Any)                                       %
%                                                                       %
%%%%%%%%%%%%%%%%%%%%%%%%%%%%%%%%%%%%%%%%%%%%%%%%%%%%%%%%%%%%%%%%%%%%%%%%%

\listoffigures

%%%%%%%%%%%%%%%%%%%%%%%%%%%%%%%%%%%%%%%%%%%%%%%%%%%%%%%%%%%%%%%%%%%%%%%%%
%                                                                       %
%     6) LIST OF TABLES (If Any)
%                                                                       %
%%%%%%%%%%%%%%%%%%%%%%%%%%%%%%%%%%%%%%%%%%%%%%%%%%%%%%%%%%%%%%%%%%%%%%%%%

\listoftables

\newpage
\listofpublications

This is a list of publications in my MPhil period. The authors come in alphabetical order.

\begin{enumerate}
    \item J. Ballweg, A.K. Goharshady, \textbf{Z. Lin} \textbf{\uline{Fast and Gas-efficient Private Sealed-bid Auctions}} In \textit{44th ACM Symposium on Principles of Distributed Computing} \textbf{(PODC)}, 2025.
    \item T. Barakbayeva, S. Farokhnia, A. K. Goharshady, P. Li, \textbf{Z. Lin} \textbf{\uline{Improved Gas Optimization of Smart Contracts.}} In \textit{11th International Conference on Fundamentals of Software Engineering} \textbf{(FSEN)}, 2025, pp. 1-10.
    \item A. K. Goharshady and \textbf{Z. Lin} \textbf{\uline{Blind Vote: Economical and Secret Blockchain-Based Voting}} In \textit{7th IEEE International Conference on Blockchain} \textbf{(Blockchain)}, 2024, pp. 46-53.
\end{enumerate}

%%%%%%%%%%%%%%%%%%%%%%%%%%%%%%%%%%%%%%%%%%%%%%%%%%%%%%%%%%%%%%%%%%%%%%%%%
%                                                                       %
%     7) ABSTRACT                                                       %
%                                                                       %
% \abstract and \endabstract are used to define a short Abstract for    %
% the Thesis.                                                           %
%                                                                       %
%%%%%%%%%%%%%%%%%%%%%%%%%%%%%%%%%%%%%%%%%%%%%%%%%%%%%%%%%%%%%%%%%%%%%%%%%

\abstract
% In decentralized blockchain environments, auctions must ensure fairness, trustless execution and bid confidentiality while operating efficiently within resource-constrained smart contracts. We propose a new family of algorithms for private, trustless auctions that protect bidder identities and bid values while remaining practical for smart contract execution. Our approach builds on top of the Dutch auction model and a stepwise revelation tree. Bidders commit to their bids using cryptographic commitment schemes and later confirm their honest following of the protocol through zero-knowledge proofs, ensuring that only the highest bid is disclosed while all other bids remain (probabilistically) confidential. A key innovation is the use of a reveal tree, which structures the bidding process into logarithmically many rounds, reducing the total number of messages to $O(\lg n)$ and thereby significantly lowering gas costs and execution times. We further explore variants introducing fake bids to optimize our probabilistic privacy guarantees.

% We ensure trustlessness by running the auction logic in a smart contract, thereby eliminating reliance on any single trusted party. This approach prevents bid tampering, front-running, and collusion by enforcing immutability and decentralized verification of bids. The resulting protocol uniquely combines efficiency, trustlessness, and enduring bid privacy, offering a scalable and secure solution for blockchain-based marketplaces and other decentralized applications.

Programmable blockchains have long been a hot research topic given their tremendous use in decentralized applications. Smart contracts, using blockchains as their underlying technology, inherit the desired properties such as verifiability, immutability, and transparency, which make it a great suit in trustless environments.

In this thesis, we consider several decentralized protocols to be built on blockchains, specifically using smart contracts on Ethereum. We used algorithmic and cryptographic tools in our implementations to further improve the level of security and efficiency beyond the state-of-the-art works. We proposed a new approach called Blind Vote, which is an untraceable, secure, efficient, secrecy-preserving, and fully on-chain electronic voting protocol based on the well-known concept of Chaum's blind signatures. We illustrate that our approach achieves the same security guarantees as previous methods such as Tornado Vote~\cite{muth_tornado_vote}, while consuming significantly less gas. Thus, we provide a cheaper and considerably more gas-efficient alternative for anonymous blockchain-based voting. On the other hand, we propose a new family of algorithms for private, trustless auctions that protect bidder identities and bid values while remaining practical for smart contract execution. We ensure trustlessness by running the auction logic in a smart contract, thereby eliminating reliance on any single trusted party. This approach prevents bid tampering, front-running, and collusion by enforcing immutability and decentralized verification of bids. The resulting protocol uniquely combines efficiency, trustlessness, and enduring bid privacy, offering a scalable and secure solution for blockchain-based marketplaces and other decentralized applications.

\endabstract

%%%%%%%%%%%%%%%%%%%%%%%%%%%%%%%%%%%%%%%%%%%%%%%%%%%%%%%%%%%%%%%%%%%%%%%%%
%                                                                       %
%     8) The Actual Contents                                            %
%                                                                       %
% The command \chapters MUST BE USED to ensure that the entire content  %
% of the Thesis is double-spaced (in version 1.0).                      %
%                                                                       %
% However, in version 2.0, \chapters will be automatically added in     %
% the beginning of the first chapter.                                   %
%                                                                       %
%%%%%%%%%%%%%%%%%%%%%%%%%%%%%%%%%%%%%%%%%%%%%%%%%%%%%%%%%%%%%%%%%%%%%%%%%

%%\chapters         % Not necessary with ustthesis.cls (v2.0).

%%%%%%%%%%%%%%%%%%%%%%%%%%%%%%%%%%%%%%%%%%%%%%%%%%%%%%%%%%%%%%%%%%%%%%%%%
%                                                                       %
% Each chapter is defined via the \chapter command. The usual sectional %
% commands of LaTeX are also available.                                 %
%                                                                       %
%%%%%%%%%%%%%%%%%%%%%%%%%%%%%%%%%%%%%%%%%%%%%%%%%%%%%%%%%%%%%%%%%%%%%%%%%

\chapter{Introduction}
\label{chp_intro}
\begin{figure}
    \centering
    \includegraphics[scale=0.8]{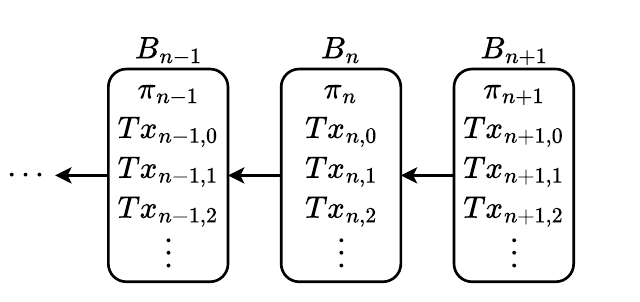}
    \caption{A simplified view of a blockchain when the block $B_{n+1}$ is appended.}
    \label{fig:blockchain}
\end{figure}

\paragraph{Blockchain}  Blockchain is a family of distributed consensus protocols, first designed by Satoshi Nakomoto as the underlying protocol of Bitcoin~\cite{nakamoto2008bitcoin}. In such protocols, our goal is to reach a consensus about an ordered sequence of \emph{transactions}. In Bitcoin, a transaction encodes transfers of money. When a user creates a new transaction, they broadcast it to the whole network using a gossip protocol. Every node on the network keeps track of the transactions they have heard of (called the \emph{mempool}) but does not consider them finalized until they are added to the \emph{blockchain}. A blockchain, as its name suggests, is a chain (singly-linked list) of blocks, with each block $B_i$ containing a sequence of transactions $\langle Tx_{i, 0}, Tx_{i, 1}, \ldots \rangle$ and a pointer to the previous block $B_{i-1}$. See Figure~\ref{fig:blockchain}. Every node keeps track of a copy of the blockchain. To ensure consensus, appending new blocks to the end of the chain is a costly endeavor, called \emph{mining}. Suppose the blockchain contains $n$ blocks. A \emph{miner} is a node that gathers unfinalized transactions, bundles them in a block $B_{n+1}$ and attempts to append this block to the end of the consensus blockchain. The block $B_{n+1}$ should also contain a proof $\pi_{n+1}$ certifying that the miner is permitted by the protocol to add this block. In Bitcoin, one needs to solve a hard proof-of-work puzzle which is based on inverting a hash function. When the puzzle is solved successfully, the miner broadcasts their block $B_{n+1}.$ The solution to the hash inversion puzzle serves as $\pi_{n+1}$. Every node then verifies the solution and adds the block to their copy of the blockchain. See~\cite{laurence2019introduction} for a more detailed treatment. Proof-of-work is not the only consensus mechanism. There are many other well-established mechanisms~\cite{DBLP:conf/crypto/DziembowskiFKP15,DBLP:conf/sac/ChatterjeeGP19,DBLP:conf/sosp/GiladHMVZ17}, such as proof-of-stake~\cite{DBLP:conf/eurocrypt/DavidGKR18} in which a miner's chance of being permitted to add the next block is proportional to their holdings in the currency.

\paragraph{Programmable Blockchains and Smart Contracts} While Bitcoin was the first cryptocurrency based on a blockchain protocol, Ethereum~\cite{wood2014ethereum} pioneered the concept of smart contracts. A smart contract is a program that is stored on the blockchain. Every node on the network keeps track of the state of every contract. This is achieved by extending what a {transaction} can do. In programmable blockchains, a transaction can (i)~transfer money, (ii)~deploy a new smart contract, providing its code -- which will be saved on the blockchain as part of the transaction, or (iii)~call a function in a previously-deployed smart contract, providing the arguments necessary for the function call.

\paragraph{Consensus} The blockchain protocol provides consensus on the history and order of transactions. Thus, every node on the network has the same view of the smart contracts, i.e.~sees the same codes deployed by transactions on the blockchain and sees the same function calls to each contract in the same order. Therefore, each node can execute the transactions in the order they appear on the blockchain and reach consensus about the state of every contract, e.g.~values of the variables internal to the contracts. This, together with the fact that smart contracts can receive and hold money in the form of the base cryptocurrency, allows one to implement real-world financial contracts as smart contracts. Of course, the underlying programming language should be unambiguous and deterministic, ensuring that different nodes executing the same sequence of function calls over the same contracts reach the same results. To achieve this, Ethereum designed a virtual machine (EVM) that supports a completely-specified low-level assembly-like bytecode format for writing smart contracts~\cite{wood2014ethereum}. The EVM bytecode language is Turing-complete~\cite{wood2014ethereum}. In practice, developers write their smart contracts in high-level languages, such as Solidity~\cite{solidity}, and then a compiler, such as \texttt{solc}, compiles it to EVM bytecode.

\paragraph{Proof of Work} Proof of Work (PoW) is used as the main consensus algorithm in Bitcoin and many other blockchain networks~\cite{nakamoto2008bitcoin} to select miners to add blocks of transactions to the network. In PoW, miners compete to solve a challenging computational problem, for example, computing a random hash value, which the best known algorithm to tackle it is by brute force. This process is called `mining'. Upon successful mining of a block, the miners will obtain a base reward and transaction fees, which serve as incentives to the miners of participating in the process~\cite{alambardar2025optimal, barakbayeva2024pixiu}. It is known that PoW suffers from the 51\% attack, meaning that the security guarantee can be breached if a malicious party gains 51\% of the total computational power~\cite{sriman2021blockchain}. PoW is also known to be highly environmentally unfriendly~\cite{chatterjee2019hybrid}. 

\paragraph{Proof of Stake}
Proof of Stake (PoS) is proposed as a consensus mechanism that addresses the problems of PoW systems - high energy consumption and scalability issues. In PoS, blocks are created by validators that are randomly chosen based on the amount of their ``staked'' cryptocurrency or collaterals they have locked up, rather than competing through computational power~\cite{DBLP:conf/eurocrypt/DavidGKR18}. This addresses the problem of energy consumption as it eradicates the intensive mining operations. Since validators staked their assets, they are incentivized to act honestly, otherwise they risk losing their collaterals. Mechanisms like slashing conditions are often implemented to penalize malicious behavior~\cite{DBLP:conf/sosp/GiladHMVZ17}. The security of PoS is hence based on economic incentives that validators with larger stakes will have stronger motivation to protect the blockchain network integrity.

\paragraph{Gas}
Since our language is Turing-complete, there is nothing to stop programmers from writing long-executing or even non-terminating contracts or contracts that use a lot of storage. As the simplest example, one can write an infinite loop \texttt{\textcolor{blue}{while}(true) \{\ldots\}} in a smart contract, deploy it on the blockchain, and then create a transaction that invokes it. In such a scenario, when this invocation is added to the blockchain, every node on the network will have to execute it, causing a deadlock. To avoid situations like this, Ethereum introduced the concept of \emph{gas}. Put simply, a gas cost is associated to every bytecode operation code (opcode). The gas cost is meant to be proportional to the actual cost of executing the operation. The costs have fixed formulas and are provided as a table in the Ethereum Yellowpaper~\cite{wood2014ethereum}. When a user creates a transaction that calls a function, they have to pay for the total gas usage of its execution, i.e.~the sum of gas costs of all invoked opcodes.  More specifically, the user has to specify the maximum amount $g$ of gas that may be used in their function call and the amount of money $p$ (in Ether) they are willing to pay per unit of gas. The transaction will first take a deposit of $g \cdot p$ from the user and then start executing the desired function call. If the transaction runs out of gas, i.e.~the invocation requires more than $g$ units of gas, it will be reverted without refunding the deposit. Otherwise, if it uses $\overline{g} \leq g$ units of gas, the user pays $\overline{g} \cdot p$ to the miners as a transaction fee and the rest is refunded~\cite{wood2014ethereum}.

Gas fees are significant and cost Ethereum users almost 4 billion USD per year~\cite{gasused}. Thus, when designing a blockchain-based protocol, we must distinguish between off-chain computation, i.e.~computation done on the user's own machine which does not incur gas costs, and on-chain computations, i.e.~calls to smart contract functions which cause computations performed by everyone on the network and do incur gas fees.

\paragraph{Commitment Scheme}
Commitment schemes are a standard cryptographic primitive and often used in blockchain-based protocols. They help mimic simultaneous actions by a group of participants. More precisely, consider $n$ participants who should each send a message to a smart contract. In the protocol, instead of directly sending a message $m$, a participant will first hash it with a nonce $r$ to produce $h=\text{hash}(m,r)$. Then, he sends the hash $h$ to the smart contract in the commit phase. The contract records the hash. In the reveal phase, each participant will send $(m, r)$ to the contract, who can in turn compute their hash and ensure the message was not changed.

The simultaneous effect is achieved because hashes in the commit phase leak no information, and hence no one can submit their messages based on any information about the other participants. Moreover, since cryptographic hash functions are collision resistant, one cannot change the message after the commitment phase.

\paragraph{Organization}
Our main contributions include the implementation and design of two protocols that are efficient and secure:

In Chapter~\ref{voting}, we present a novel blockchain-based voting protocol using a combination of commitment schemes and Chaum's blind signatures as our cryptographic primitives. Our protocol provides the same security guarantees as previous methods, such as Tornado Vote~\cite{muth_tornado_vote}. However, it is important to note that, unlike previous approaches, we purposely avoid zero-knowledge proofs and zkSNARKS. This is an intentional choice to reduce the gas usage (execution costs) of the resulting smart contract. Thus, we present a cheap and gas-efficient anonymous blockchain-based voting protocol without compromising any of the usual security guarantees.

In Chapter~\ref{auction}, we provide a novel and simple auction protocol that can be implemented as a smart contract and combines ideas from Dutch auctions, commit-reveal schemes and binary interval trees. Our protocol takes $O(\lg m)$ time where $m$ is the maximum allowed bid. Similarly, every bidder in the protocol pays for $O(\lg m)$ units of gas in the worst case. Our protocol is decentralized, premissionless, and trustless. It also guarantees both bid independence, i.e.~every bidder is unaware of others' bids when making their own, and privacy for losing bidders, ensuring that only the highest bid and its corresponding bidder are publicly identified. Finally, we guarantee that the results are publicly verifiable.

\newpage

\chapter{Preliminaries}
\label{chp_preliminaries}
% \section{Commitment Scheme}
% Commitment schemes are a standard cryptographic primitive and often used in blockchain-based protocols. They help mimic simultaneous actions by a group of participants. More precisely, consider $n$ participants who should each send a message to a smart contract. In the protocol, instead of directly sending a message $m$, a participant will first hash it with a nonce $r$ to produce $h=\text{hash}(m,r)$. Then, he sends the hash $h$ to the smart contract in the commit phase. The contract records the hash. In the reveal phase, each participant will send $(m, r)$ to the contract, who can in turn compute their hash and ensure the message was not changed.
 
% The simultaneous effect is achieved because hashes in the commit phase leak no information, and hence no one can submit their messages based on any information about the other participants. Moreover, since cryptographic hash functions are collision resistant, one cannot change the message after the commitment phase.

\section{Blind Signatures}
The concept of a blind signature was first introduced by Chaum in \cite{chaum_blind_signatures} to provide both anonymity and privacy to payees in digital cash systems. It allows a payer to obtain a certificate from the bank that blinds it so that the bank can only know the proof of payments but not the identities of the payers. Unsurprisingly, this has already been used for voting, too~\cite{schmid_blind_signatures}. However, both the concept of blind signatures and the voting protocols building upon it predate blockchains. 

In Chapter~\ref{voting}, we use the most classical implementation of blind signatures using RSA~\cite{rivest_a_method}.  Suppose the bank has an RSA public key $(N,e)$ and its corresponding private key $d,$ and that  Alice wants to pay Bob 1 dollar. The protocol goes as follows:
\begin{itemize}
	\item Alice constructs a banknote, which is a string $b =$ `This is a banknote with serial XXX...XXX'. The serial number is a random value chosen by Alice. She then computes $h = \text{hash}(b)$ using a pre-defined cryptographic hash function.
	\item Alice chooses a random number $r$ and keeps it secret. She computes $h'=h \cdot r^e$ and sends it to the bank. Note that, as standard in RSA, all calculations are done modulo $N$ and $r^e$ is the result of encrypting $r$ using the bank's public key $e.$
	\item The bank signs $h'$ and sends $h'^d$ to Alice. It also deducts 1 dollar from Alice's balance.
	\item Knowing $r$, Alice can easily compute its modular multiplicative inverse $r^{-1}.$ She then obtains the bank's signature on $h,$ i.e.~$h^d,$ by a simple calculation: $$h'^d \cdot r^{-1}=h^d \cdot r^{e \cdot d} \cdot r^{-1} = h^d \cdot r \cdot r^{-1} = h^d.$$
	
	\item Alice sends $(b, h^d)$ to Bob.
	\item Bob immediately sends $(b, h^d)$ to the bank. The bank checks that $h^d$ is a correct signature on the hash $h = \text{hash}(b).$ It also checks that $b$ is well-formed and the serial number in $b$ has not been used before. If the checks pass, it increases Bob's account balance by 1 dollar.
\end{itemize}

The beauty of the protocol above is that the bank never saw $b$ or $h$ when signing $h'.$ Indeed, knowledge of $h' = h \cdot r^e$ does not give the bank any information about $h$ due to the existence of the random nonce $r,$ which serves as the \emph{blinding factor}. Thus, when presented with $(b, h^d)$ by Bob, the bank is able to verify that $b$ is indeed a valid banknote signed by itself at some point, but it cannot unmask Alice or distinguish her or her banknote from any other banknote of the same denomination.

In Chapter~\ref{voting}, we will develop the idea of using blind signatures to mask voters' identities so as to break the link between the voter and their vote and thus achieve secrecy.

\section{Voting}

\paragraph{Traditional Voting}
Voting is a democratic process that requires both confidentiality and accountability. In a typical voting scenario, every participant or third party should be able to verify the result, i.e.~the tally, of the process but no participant's choice shall be leaked. In physical voting protocols, voters have to show up in person to cast their ballots and are only informed of the results after a centralized organization performs tallying. Of course, if the voting is for an office, the candidates will each have representatives in the tallying process to create more trust in the system. Nevertheless, this process is opaque and effectively a black box from the point-of-view of the voters and hence undermines voter confidence. See~\cite{moura_blockchain_voting} for a more detailed discussion of this point.

\paragraph{Electronic Voting}
Designing schemes and protocols for electronic voting has been a hot research topic for several decades. We refer to~\cite{mursi2013development} for an excellent survey. In this work, we are especially interested in blockchain-based voting. This is because smart contracts hosted on the blockchain effectively inherit many of its characteristics, such as verifiability and transparency, which are desirable in a voting protocol. The literature in this domain is vast and there is no way we can do justice to all the previous approaches. Thus, we refer to~\cite{hjalmarsson_blockchain_based} for a survey of blockchain-based voting methods. We cover some of the most related previous works in Section~\ref{voting_previous_work}. Specifically, the closest previous work is Tornado Vote~\cite{muth_tornado_vote} which provides an anonymous blockchain-based voting protocol based on zero-knowledge proofs. 

\section{Auctions}

\paragraph{Transparent Auctions}
An auction protocol is called \emph{transparent} if every participant's bid is publicly revealed by the end of the protocol. A common approach in transparent auctions is the commit-reveal scheme, as implemented in several smart contract-based protocols such as~\cite{pop2020ethereum}. In these schemes, bidders first commit to their bids and then reveal them during a designated phase. Although this method ensures trustless execution, it ultimately exposes all bid values, offering no bid privacy. Unlike traditional commit-reveal auctions, we aim to ensure privacy for the losing bidders, i.e.~guarantee that no information is leaked about their bids.

\paragraph{Anonymous Auctions}
An auction protocol is \emph{anonymous} if it leaks the bids but no one can infer their ownership and know which bid belongs to which bidder. Two prominent directions in designing anonymous auctions are based on ring signatures and blind signatures.

\paragraph{Ring Signatures}
A ring signature is a cryptographic primitive that allows any member of a group to sign a message anonymously, making it infeasible to determine which member produced the signature. It has been proposed for building anonymous auction systems where the bid values can be seen by everyone after the auction ends but they are not tied to someone's identity. For example,~\cite{ye2023anonymous} employs ring signatures in a blockchain setting to achieve bidder anonymity. However, the approach is not fully trustless since the auctioneer may later deanonymize the bidders. Similarly,~\cite{huang2021ba2p} implements an anonymous first-price sealed-bid auction using ring signatures, relying on a centralized auctioneer with the potential to compromise anonymity. In~\cite{sharma2021}, ring signatures are used to protect bid confidentiality and bidder identity, yet the auctioneer still retains the capability to reveal bid values.
 
\paragraph{Private Auctions}
We say an auction is private if it does not leak any information about the losing bids. See Section~\ref{sec:setting} for a formal definition. Most private auctions in the literature are based on either homomorphic encryption or multiparty computation protocols. They are not designed for the blockchain setting and often require costly computations that, if implemented in a smart contract, would lead to an untenable gas usage of $\Omega(n)$ per participant, where $n$ is the number of bidders.

\paragraph{Multiparty Computation}
Many auction protocols such as~\cite{yao1982protocols, ghasaei2023privateauction, zhang2024blockchain, montenegro2013mobilenetworks, novakovic2024cryptobazaar} utilize secure multiparty computation to secretly compute the result of an auction while keeping the bids private. For example, Cryptobazaar~\cite{novakovic2024cryptobazaar} is a protocol that runs in $O(n)$ time and can be generalized to an $i$th-price auction that discloses only the $i$th highest bid and nothing else. It uses the Anonymous Veto protocol of \cite{hao20092round} to blind the bids from the auctioneer and everyone else.

% \paragraph{Blind Signatures} Blind signatures~\cite{DBLP:conf/crypto/Chaum82} are digital signatures in which the message is first ``blinded'' so that the signer cannot see its content, ensuring that the signature cannot later be linked back to the original message. This method is used in \cite{xiong2019auction}, where blind signatures combined with time-lock encryption decouple bid values from bidder identities. Although the bid values become publicly verifiable after the auction ends, they are not tied to individual bidders. 

\paragraph{Homomorphic Encryption}
Homomorphic encryption is a cryptographic tool that allows users to compute directly on encrypted data without having to decrypt it first. It naturally fits well in auction protocols, as bidders may securely perform calculations on their encrypted bids~\cite{DBLP:conf/crypto/MalavoltaT19,DBLP:conf/fc/BogetoftDJNPT06}. For example, a Pedersen commitment \cite{pedersen1992non}, which is a type of homomorphic commitment scheme, is deployed in \cite{tyagi2023riggs} and \cite{glaeser2023cicada}. In the Riggs protocol~\cite{tyagi2023riggs}, each bidder has a balance (commitment) recorded in the auction house, which represents the total amount each bidder may use to place bids in the auctions being hosted. A Pedersen commitment is particularly useful in this case as bidders can directly update their balances or place a bid without revealing the amounts.
\newpage

\chapter{Blockchain-based Voting}
\label{voting}
This chapter is based on the following publication:

\begin{enumerate}
    \item A. K. Goharshady and \textbf{Z. Lin} \textbf{\uline{Blind Vote: Economical and Secret Blockchain-Based Voting}} In \textit{7th IEEE International Conference on Blockchain} \textbf{(Blockchain)}, 2024, pp. 46-53.
\end{enumerate}
\newpage

In this chapter, we introduce Blind Vote, which is based on Chaum's blind signatures as our cryptographic primitive. We deliberately eliminate the use of zero-knowledge proofs and zkSNARKS to reduce gas usage in the smart contract. The resulting protocol achieves the same or even higher security guarantees as previous works.

\section{Related Works} \label{voting_previous_work}

Electronic voting is a vast field with many contributions. Since it would be impossible to enumerate all of the many approaches developed over decades of research, in this section, we consider several of the most related previous works. We refer to~\cite{mursi2013development,hjalmarsson_blockchain_based,moura_blockchain_voting} for a more detailed overview of other voting methods.

\paragraph{Desired Properties of an Electronic Voting Protocol} The early work \cite{fujioka_practical}, which predates blockchain, identifies the required security properties of a secure electronic voting system as follows (quoted from~\cite{fujioka_practical}):
\begin{itemize}
	\item Completeness: All valid votes are correctly counted.
	\item Soundness: The voting cannot be disrupted by any single malicious voter.
	\item Verifiability: The result of the voting cannot be falsified by anyone.
	\item Unreusability: Each voter can vote only once.
	\item Privacy: All votes remain secret to other party.
	\item Fairness: The voting cannot be affected by anything.
	\item Eligibility: Only voters that are allowed to vote can vote.
\end{itemize}

\paragraph{Overview of Previous Works}
Many of these properties are attained by default when the voting is implemented as a smart contract. Most importantly, if anyone is eligible to vote, then privacy is achieved by default on blockchain since one cannot associate an account, which is basically a public-private key pair, to a real-life person. However, deanonymization and profiling can still pose threats to privacy~\cite{beres_blockchain}. Additionally, in the natural case that we have a predefined set of eligible voters identified by their public keys (accounts), privacy is no longer a given since every transaction on the blockchain is public and traceable. Therefore, \emph{untraceability}, i.e.~a disconnect between a voter's public key and their vote, is also needed for a protocol to be secure. We often use the word \emph{secrecy} as a shorthand for untraceability and privacy. There are a wide variety of protocols that aim to achieve secrecy. Examples include the use of homomorphic encryption~\cite{Anggriane_Advanced}, anonymous off-chain communication channels~\cite{fujioka_practical,carcia_blockchain_based} and, most commonly, standardized tokens and zero-knowledge proofs \cite{mccorry_boardroom,hao_anonymous_voting,muth_tornado_vote,killer_provotuMN,killer_eternum}. 

Some approaches sacrifice secrecy or provide a weaker guarantee of privacy. For example, in ~\cite{fujioka_practical} voters first send their votes to an \textit{administrator} for it to add a signature using blinding techniques. After retrieving the signatures, the voters then forward the votes to a \textit{counter} for it to count the votes and accumulate the results. 
Although being scalable, this protocol uses an anonymous communication channel as a means to break the link between voters and their votes. However, this practice has two drawbacks: (i)~the counter is centralized, and (ii)~in the absence of the blockchain protocol, the voters have to perform off-chain communications with the administrator and the counter. These communications can potentially be traced by internet service providers or other intermediaries and used to unmask the voters~\cite{ayed_conceptual_secure}. Moreover, completeness can be violated if the centralized entities refuse to process valid communications from a voter.

\paragraph{Secrecy via Homomorphic Encryption}
The work \cite{Anggriane_Advanced} presents a voting protocol that uses a cryptosystem with an additive homomorphic property to achieve anonymity. The idea is pretty elegant. Here, we provide a simplified outline. In the Paillier cryptosystem, for any two messages $m_1$ and $m_2,$ we can multiply (aggregate) their encryptions to obtain an encryption of $m_1 + m_2.$ This can be adapted to voting in a natural way. An administrator first publishes their Paillier public key on the smart contract. The voters then encrypt their votes using this public key before submitting them to the contract. The contract tallies the votes by multiplying ciphertexts. When the voting is over, the administrator decrypts the final (tallied) ciphertext and hence reveals the final results. 

Although the ciphertexts (encrypted votes) are visible all the time on the blockchain, one cannot decrypt them without the private key and hence the votes are secret from the network's point-of-view. However, a lethal drawback of this scheme is that there is always an administrator who should set up the voting and hence possesses the private key. They can always decrypt the ciphertexts off-chain. Thus, there is no secrecy against the administrator. Also, a voter's vote cannot be verified efficiently as it should be encrypted. So, a malicious voter can cast an invalid ballot and affect the overall correctness of the results.

\paragraph{Tornado Vote}
Tornado Vote~\cite{muth_tornado_vote} is the most recent and one of the closest related works. It achieves all the desired properties listed above. At its core, Tornado Vote uses a cryptocurrency mixer called Tornado Cash~\cite{pertsev2019tornado} together with zero-knowledge proofs and a relayer infrastructure to achieve secrecy. It uses its own custom ERC-20 for each election. The basic idea is to use zero-knowledge proofs to disconnect voter identities from their votes. 

Tornado Vote considers three types of users: administrator, voter and relayer. The role of the administrator is to set up the voting and give eligibility tokens to voters. The role of the relayers, who are often accessed through a secure channel such as Tor, is to send messages to the smart contract on behalf of the voters, ensuring that the source of a message cannot be identified. Note that our protocol does not rely on Tor and obtaining privacy between the users and relayers using Tor, VPNs or other tools is an orthogonal problem. Since each vote is effectively a token (a piece of currency), voting between $k$ options can be seen as a transfer of money from the voters to one of $k$ predetermined accounts. Mixers, such as Tornado Cash~\cite{pertsev2019tornado}, enable such transfers in a way that disconnects the sender and recipient. In a voting setting, this mixing property is exactly the same as the secrecy property, i.e.~the sender is the voter and the target account is the chosen vote. 

Despite providing all the desired security guarantees, a major drawback of Tornado Vote is its high gas usage. Indeed, the authors make several gas-optimizing choices, such as using different hash functions in various stages, to ameliorate this problem~\cite{muth_tornado_vote}. However, the issue is inherent and already present in Tornado Cash. Its root cause is the necessity of sending zero-knowledge proofs to the smart contract and verifying them on-chain.

In this work, we provide an alternative method which does not require zero-knowledge proofs at all and instead builds upon much more gas-efficient cryptographic primitives such as blind signatures and basic commitment schemes. The result is a huge saving in the overall gas costs of the election. Voting methods based on blind signatures were previously considered in~\cite{DBLP:journals/iacr/LiuW17,DBLP:conf/ithings/HardwickGAM18}. In comparison with these protocols, our approach (and TornadoVote) provide stronger privacy guarantees, as well as the added ability to delegate votes to third-parties.

\section{Blockchain-based E-voting protocol using Blind Signatures}
In this section, we describe our protocol for blockchain-based secret voting using blind signatures. As in Tornado Vote~\cite{muth_tornado_vote}, our approach also considers three types of users: an administrator, $n$ voters and a number of relayers.

\paragraph{Step 0. Deployment} The administrator deploys the Blind Vote contract on the blockchain. During deployment, the following values are set in the contract's constructor (chosen by the administrator):
\begin{itemize}
	\item The maximum number $n_{\texttt{max}}$ of allowed voters.
	\item A registration fee $f$ that has to be paid by every voter;
	\item A relay reward $\rho$ that will be paid to each relay;
	\item A deposit $\delta,$ which is paid at the time of deployment by the administrator;
	\item Time limits $t_1 < t_2 < \ldots < t_6 $ for the following steps of the protocol. Smart contract functions in each step $j$ of the protocol can only be called after time $t_{j-1}$ and before or on $t_j.$
\end{itemize}
The administrator has to ensure that $\rho$ is large enough to cover the gas fees for relays and additionally incentivize them, and that $f$ and $\delta$ are large enough for the contract to be able to pay all relays.

\begin{figure}[H]
	\begin{center}
	\includegraphics[width=.5\linewidth]{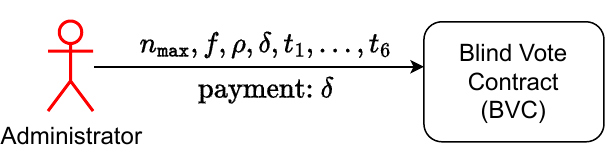}
	\end{center}
	\caption{An illustration of Step 0 of Blind Vote.}
\end{figure}

\begin{figure}[H]
	\begin{center}
	\includegraphics[width=.6\linewidth]{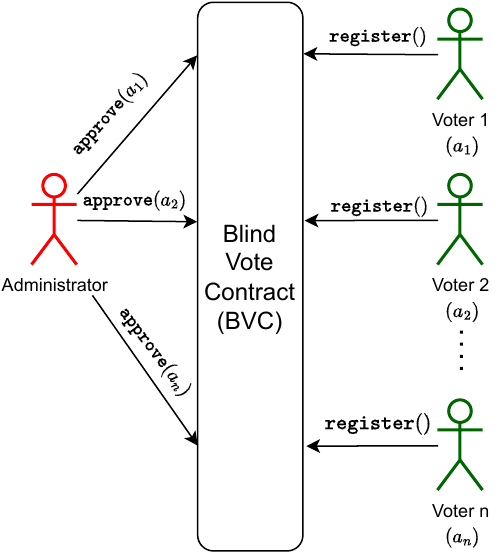}
	\end{center}
	\caption{An illustration of Step 1 of Blind Vote.}
    \label{fig:voting-step1}
\end{figure}

\paragraph{Step 1. Voter Registration} This step is open until time $t_1$. For every eligible voter $i$ who has address $a_i$ on the blockchain, the administrator calls a function $\texttt{approve}(a_i),$ adding the voter's address to the voting roll. The contract keeps track of all $a_i$'s. Additionally, each voter should register in the same step, i.e.~by time $t_1,$ by calling the $\texttt{register}()$ function in our smart contract and paying a deposit of $f.$ A voter can take part in the remainder of the protocol only if both the registration and approval are done by time $t_1.$ A voter who registers but is not approved by time $t_1$ can receive a refund after $t_1$ by calling $\texttt{step1\_refund}()$. The contract keeps track of the total number $n$ of valid voters and their addresses and would not allow $n$ to exceed the maximum set in the previous step. As shown in Figure~\ref{fig:voting-step1}, we use the color red for the administrator and green for voters.

\begin{figure}[H]
    \begin{center}
        \includegraphics[width=.6\linewidth]{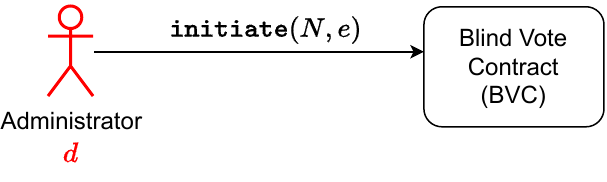}
    \end{center}
    \caption{An illustration of Step 2 of Blind Vote.}
    \label{fig:voting-step2}
\end{figure}

\paragraph{Step 2. Initialization} The administrator generates an RSA public key $(N, e)$ and the corresponding secret key $d$. He calls the function $\texttt{initiate}(N, e)$ of the contract. The contract records $N$ and $e,$ which are now public knowledge. Here, $N$ is the RSA modulus and $x^e \mod N$ is the encryption/signature verification of $x$. Conversely, $y^d \mod N$ is the decryption/signature on $y$. As shown in Figure~\ref{fig:voting-step2}, we show secrets in red and public information in black.

\paragraph{Step 3. Delegation} Each voter $i$ chooses an RSA public key $(N_i, e_i)$ and a corresponding secret key $d_i.$ She keeps all of these values secret for the moment. The voter's goal is to delegate her voting rights to anyone who can sign using $d_i,$ i.e.~herself, while making sure that no one can connect $d_i$ to her publicly-known blockchain address $a_i.$ For this, she uses a blind signature scheme as follows:
\begin{itemize}
    \item Compute $h_i = \text{hash}(N_i, e_i).$
    \item Choose a random blinding factor $r_i$ and calculate $h'_i = h_i \cdot r_i^e \mod N.$ Recall $e$ is administrator's public key.
    \item Submit $h'_i$ to the contract by calling $\texttt{delegate}(h'_i).$ The contract records the value of $h'_i.$
\end{itemize}
The goal is to get the administrator's signature on $h_i,$ i.e.~$s_i := h_i^d \mod N$, which serves as a proof that anyone controlling the private key corresponding to $(N_i, e_i)$ can cast a vote.

\begin{figure}[H]
    \begin{center}
        \includegraphics[width=.6\linewidth]{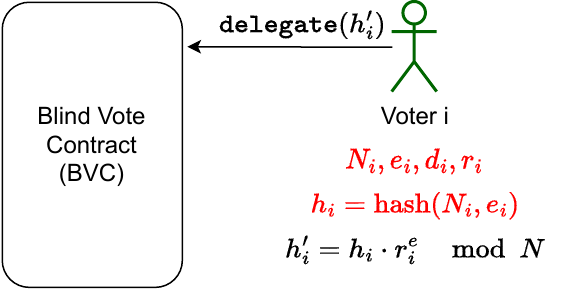}
    \end{center}
    \caption{An illustration of Step 3 of Blind Vote.}
\end{figure}

\paragraph{Step 4. Blind Signature} For every $h'_i$ provided in the previous step, the administrator computes the signature $s'_i = {h'_i}^d \mod N$ and announces it to the contract by calling $\texttt{blind\_sign}(a_i,  s'_i).$ The contract checks that $s'_i$ is a valid signature on $h'_i$ and, if so, stores it. The voter $i$ can now unblind the signature on her own machine by computing
$$
s_i = s'_i \cdot r_i^{-1}  = {h'_i}^d \cdot r_i^{-1} = h_i^d \cdot r_i^{e \cdot d} \cdot r_i^{-1} = h_i^d \cdot r_i \cdot r_i^{-1} = h_i^d,
$$
where all calculations are done modulo $N.$
The latter is the administrator's RSA signature on $h_i = \text{hash}(N_i, e_i).$ Thus, the voter now has the administrator's signature on her own RSA public key without having revealed it to the administrator or anyone else.

\begin{figure}[H]
    \begin{center}
        \includegraphics[width=.8\linewidth]{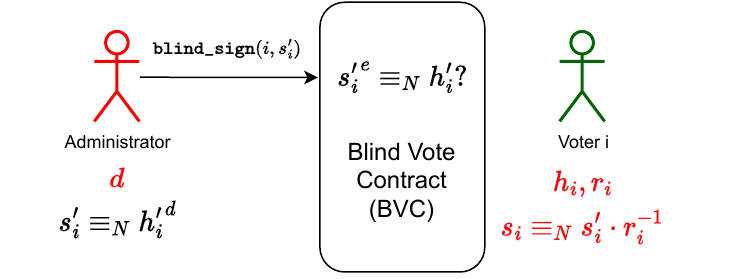}
    \end{center}
    \caption{An illustration of Step 4 of Blind Vote.}
\end{figure}

 At this point, each voter $i$'s ability to cast a vote is delegated to the RSA private key she chose and is no longer connected to her blockchain identity/account address $a_i$. Specifically, anyone who owns the secret key $d_i$ corresponding to a public key $(N_i, e_i)$ whose hash $h_i = \text{hash}(N_i, e_i)$ is signed by the administrator can cast a vote. In other words, the administrator's signature on $h_i$ is seen as a proof of eligibility for the owner of the corresponding secret key to have one vote in the election.

 \paragraph{Relaying} In the next step, voter $i$ will choose her vote $v_i.$ However, she cannot simply send this vote to the contract, since that would (i)~leak her identity, and (ii)~allow other voters to see her vote before deciding theirs. To overcome (i), we use the standard technique of \emph{relaying}. A \emph{relay} is a blockchain participant who is willing to submit a function call to the contract on behalf of a voter in exchange for a reward. As is standard, we assume that the voters can send anonymous messages to a public notice board that is seen by relays. We also assume that this does not leak their identity or IP address as they can use services such as Tor to hide this information. A relay can then check if a function call is profitable for them, and if so, is incentivized to make the call on behalf of the voter. Our relaying mechanism matches those of Tornado Vote~\cite{muth_tornado_vote} and Tornado Cash~\cite{pertsev2019tornado}. To solve problem (ii), we apply a standard commitment scheme.

\paragraph{Step 5. Commitment} Each voter $i$ who wants to vote $v_i$ chooses a random nonce $x_i$ and computes $c_i := \text{hash}(v_i, x_i).$ There is a function $\texttt{commit}(N_i, e_i, s_i, c_i, sc_i)$ in the smart contract that can be called by \emph{anyone on the blockchain network}, including relays. This function is used to commit to a particular vote. When this function is called, the contract checks the following:
\begin{itemize}
    \item The \texttt{commit()} function was previously called successfully no more than $n$ times.
    \item $(N_i, e_i)$ is a valid RSA public key.
    \item $s_i$ is the administrator's RSA signature on the hash of the public key $(N_i, e_i).$ In other words, $s_i^e = \text{hash}(N_i, e_i) \mod N.$
    \item $c_i$ is a string that serves as the commitment to a vote.
    \item $sc_i$ is a valid RSA signature on $c_i$ using the private key corresponding to $(N_i, e_i).$ Formally, $sc_i^{e_i} = c_i \mod N_i.$
    \item This is the first time this function is called and passed the checks above for the current combination of $(N_i, e_i, s_i).$
\end{itemize}
If all these checks pass, the contract records the commitment $c_i.$ It also pays a reward of $\rho$ to the caller of the \texttt{commit()} function, who is presumed to be a relay.

\begin{figure}[H]
    \begin{center}
        \includegraphics[width=.7\linewidth]{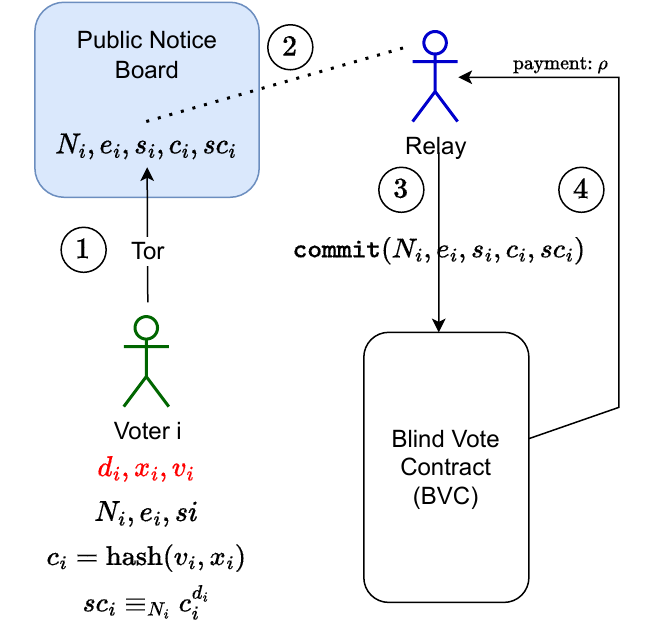}
    \end{center}
    \caption{An illustration of Step 5 of Blind Vote.}
\end{figure}

\paragraph{Step 6. Revealing} Finally, after all the commitments to the votes are submitted to the contract in the previous section, the voters reveal their votes. This is also done through a relay to preserve their privacy. Specifically, there is a function $\texttt{reveal}(c_i, v_i, x_i)$ which can be called by anyone, including the relays. This function checks the following:
\begin{itemize}
    \item $c_i$ was a commitment from the previous step and was not revealed before.
    \item $\text{hash}(v_i, x_i) = c_i.$
\end{itemize}
If the checks pass, the contract records the vote $v_i,$ updates the tally as necessary, and pays a reward of $\rho$ to the caller of the $\texttt{reveal}()$ function who can be the relay\footnote{It is possible for the voter to call this function herself if she does not care about secrecy. The same applies to $\texttt{commit}().$}.

\begin{figure}[H]
    \begin{center}
        \includegraphics[width=.7\linewidth]{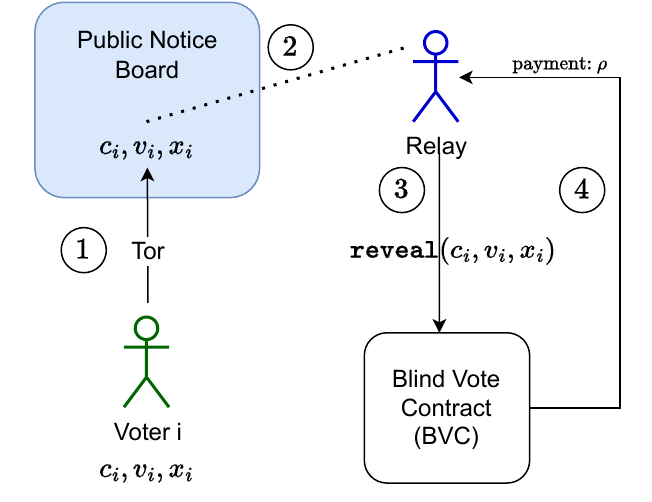}
    \end{center}
    \caption{An illustration of Step 6 of Blind Vote.}
\end{figure}

If all the steps above are performed correctly, then the votes are submitted to the smart contract using the RSA identities $(N_i, e_i)$ which are blinded and thus disconnected from the voters' actual blockchain identity/account address $a_i.$ So, we have a working blockchain-based protocol for secret voting using only blind signatures and commitment schemes.

\paragraph{Verifications, Incentives and Penalties} To ensure that all parties follow the protocol correctly, we have the following incentive structure:
\begin{itemize}
	\item After Step 1, any voter who fails to register is excluded from voting. 
	\item After Step 3, any registered voter who fails to successfully call $\texttt{delegate}()$ loses the ability to vote, but also her deposit $f.$
	\item After Step 4, if the administrator fails to sign one of the $h'_i$ values provided by a voter in the previous step, the voting is canceled and this is seen as cheating. This can be reported by anyone by calling $\texttt{report\_refused\_signature}(i).$ Thus, the administrator's deposit $\delta$ is confiscated and paid to the voters. More specifically, each voter $i$ can call a function $\texttt{step4\_refund}()$ to receive $f + \delta/n$ in her original address $a_i.$ This will deter the administrator from cheating and, assuming $d$ is chosen to be large enough, ensures that the voters are compensated for their gas fees and also get their deposits back. Thus, the administrator has to provide exactly $n$ signatures in Step 4 for the protocol to continue.
	\item In Steps 5 and 6, relays are incentivized to submit the commitment/revealing function calls on behalf of the voters since they will receive a reward of $\rho$ for this.
	\item After Step 5, if the $\texttt{commit}()$ function is successfully called more than $n$ times, this means the administrator cheated and provided extra RSA signatures in addition to the ones in Step 4. In this case, the voting is canceled again, the administrator's deposit is confiscated and paid to the voters. As before, each voter $i$ can withdraw $f + \delta'/n$ into her account $a_i$ by calling $\texttt{step5\_refund()}.$ Here, $\delta'$ is the remaining deposit of the administrator, after subtracting the relay fees.
	\item A voter who fails to submit her commitment in Step 5 has effectively failed to vote. We assume everyone is naturally incentivized to vote and no one would voluntarily decide not to commit at this step. The same applies to revealing in Step 6.
	
	\item If all steps are performed correctly and none of the cases above happen, the administrator can call $\texttt{admin\_refund}()$ after time $t_6$ to receive his deposit $d$ back. Similarly, each voter $i$ can call $\texttt{voter\_refund}()$ to receive a refund of $f - 2 \cdot \rho,$ i.e.~her initial deposit minus the relaying fees for her messages in Steps 5--6.
\end{itemize}

\paragraph{Generality of Votes} We note that blind vote can support any system of voting since we are not assuming any particular structure on the votes $v_i.$ Moreover, the tallying can follow any desired formula and the votes do not have to necessarily be a choice out of a fixed set of options. In this sense, our approach is strictly more general than Tornado Vote~\cite{muth_tornado_vote}, in which every voter has to choose a vote from a pre-fixed set of possible options. For example, our approach would allow proportional ranked choice voting~\cite{bartholdi1991single}.

\paragraph{Delegation of Voting Rights} In Blind Vote, a voter can easily delegate her voting rights to someone else. In Step 3, the voter $i$ is delegating her voting rights to anyone who has the RSA secret key $d_i$ corresponding to the public key $(N_i, e_i).$ When explaining the algorithm, we presumed that the RSA key pair is generated by the voter herself. However, if she wants to delegate the voting rights to someone else, she can ask them to generate their key pair and only give their public key to her. She will then use this public key in Steps 3 and 4, and provide the unblinded signature $s_i$ to the delegate. Knowing $s_i,$ the delegate takes over Steps 5 and 6 and votes.

\paragraph{Improvements in Gas} There are a number of ways in which we can improve the gas usage of our protocol above, mainly by reducing the amount of storage used by the contract or moving parts of the computations off-chain. We assume that the voters have a secure communication channel with the administrator. We can thus apply the following optimizations:
\begin{itemize}
    \item \emph{Moving Steps 3 and 4 off-chain.} In Step 3, each voter $i$ sends her $h'_i$ directly to the administrator. This message is authenticated and includes a signature $\sigma_i$ corresponding to the user's blockchain identity $a_i.$ The administrator then signs $h'_i$ and sends the signature $s'_i$ back to voter $i.$ This whole communication happens off-chain. Only if the administrator fails to provide a valid blind signature $s'_i$ off-chain does the voter call the $\texttt{delegate}()$ function on-chain and the administrator will then be required to call $\texttt{blind\_sign}(a_i, s'_i)$ on-chain, too. If the voter has already received a blind signature on $h'_i$ but then demands another blind signature on a different value $h''_i,$ then the administrator can call a function $\texttt{report}(i, h'_i, \sigma_i).$ This proves that the voter is trying to cheat, allowing the administrator not to provide another signature and also confiscating the voter's deposit. 
     
    This change ensures that, as long as both the voter and administrator are rational and thus prefer not to pay extra gas fees, Steps 3 and 4 can be done off-chain and for free. However, if any party tries to be dishonest, then the normal on-chain protocol will be followed and both will be required to pay gas fees (and potentially also lose their deposit). 
    
    \item \emph{Premature Commitment.} Suppose the voter has already chosen her vote $v_i$ after Step 2. We can modify the protocol and consider a variant in which the voter does not try to obtain a blind signature on her own public key $(N_i, e_i)$ in Steps 3 and 4, but instead tries to get the administrator's signature directly on her commitment $c_i = \text{hash}(v_i, x_i).$ In this variant, Step 3 will change so that we have $h_i = c_i.$ Step 5 will then be simplified with the $\texttt{commit}(s_i, c_i)$ function needing access to only $s_i$ and $c_i$ and verifying that $s_i$ is the administrator's signature on $c_i,$ i.e.~$s_i^e = c_i \mod N.$ While this idea reduces the gas usage, the tradeoff is that it precludes the possibility of delegating the voting rights to a separate person as outlined above.
\end{itemize}

\section{Security Analysis}
We now provide brief arguments as to why Blind Vote satisfies all the desired security properties of a secret voting scheme. The most important property, i.e.~secrecy, is naturally inherited from blind signatures. Most other properties are inherited directly from the blockchain.

\begin{enumerate}
    \item \emph{Eligibility:} The eligibility to vote is established in Step 1, where the administrator approves all eligible voters. This can also be moved to Step 0, by asking the administrator to provide a hard-coded list of eligible voters. No one other than the eligible voters or the administrator can compute the the blind signatures needed to make a commitment in Step 5. If the administrator cheats and adds extra commitments, there will be more than $n$ valid commitments and the contract penalizes him and cancels the vote. So, there is no incentive for such cheating.
    \item \emph{Completeness:} As long as the time allocated to each step is sufficiently long to ensure the voters/relayers will be successful in calling the contract's functions, all valid votes will be committed to in Step 5 and then revealed in Step 6. This ensures completeness.
    \item \emph{Soundness:} No voter's conduct has any effect on the other voters' ability to vote. A voter who does not follow the protocol correctly can only lose her own voting right / deposit but cannot disrupt the voting.
    \item \emph{Secrecy:} This important property is inherited from blind signatures. Since each voter's blockchain identity / account address $a_i$ is completely disconnected from the RSA keys she uses to cast her vote by a blinding process in Steps 3 and 4, there is no way to distinguish the source of a particular vote or the vote of a particular voter.
    \item \emph{Unreusability:} Every voter can obtain only one signature of the administrator on a single hash in Step 4. This signature can then be used only once to commit to a single vote in Step 5. Thus, no voter can vote twice.
    \item \emph{Fairness:} No one can affect the voting or its results. The administrator is obliged by his deposit to provide the blind signatures correctly in Step 4. As argued, he cannot add extra signatures either. Each voter votes exactly once. A relay cannot affect the results of the voting since they can only get paid their reward $\rho$ if they relay a correct message and everything is verified by the contract. An outside party other than the administrator, voters and relays, has no way of affecting the contract or calling any of its functions.
    \item \emph{Verifiability:} The blind signatures and commitment schemes are automatically verified by the smart contract and any function call that violates them is automatically rejected. However, since blockchain data is public, anyone with access to the blockchain can separately verify the correctness of the results on their own.
\end{enumerate}

\section{Implementation and Performance Analysis} \label{gas_usage}

We have implemented Blind Vote as an Ethereum smart contract written in Solidity. As mentioned above, in Blind Vote most of the computations are moved off-chain and hence do not incur gas costs. Moreover, the on-chain computations involve simple and efficient operations such as verifying RSA signatures or computing hashes. We intentionally avoided gas-inefficient operations such as on-chain verification of zero-knowledge proofs. We also store only a tiny amount of information on-chain. 
To obtain exact gas consumption numbers, we deployed our contract on Remix~\cite{remix}, allowing us to calculate the gas usage of each function call, which is shown in Table~\ref{table:function_gas_usage}.

    \begin{table}
        \begin{tabular}{C{0.13\linewidth} C{0.3\linewidth} C{0.13\linewidth} C{0.13\linewidth} C{0.17\linewidth}}
            \hline
            \hline
            Step & Function & Min & Max & Paid by \\
             \hline \hline
            \multirow{1}{0.2cm}{0} & constructor & - & 6967$k$ & Admin\\
            % \hline
            \multirow{2}{0.2cm}{1} & approve & - & 71$k$ & Admin\\
            & register & 87$k$ & 123$k$ & Voter \\
            % \hline
            \multirow{1}{0.2cm}{2} & initiate & - & 358$k$ & Admin \\
            % \hline
            \multirow{1}{0.2cm}{3} & delegate & - & 79$k$ & Voter \\
            % \hline
            \multirow{1}{0.2cm}{4} & blind\_sign & 80$k$ & 259$k$ & Admin \\
            % \hline
            \multirow{2}{0.2cm}{5} & commit & 294$k$ & 309$k$ & Relay \\
            & commit\_premature & 119$k$ & 210$k$ & Relay \\
            % \hline
            \multirow{1}{0.2cm}{6} & reveal & 94$k$ & 152$k$ & Relay \\
            \hline
            \hline
            \multirow{5}{0.7cm}{Refund} & admin\_refund & - & 52$k$ & Admin \\
            & voter\_refund & - & 83$k$ & Voter \\
            & step1\_refund & - & 86$k$ & Voter \\
            & step4\_refund & - & 62$k$ & Voter \\
            & step5\_refund & - & 76$k$ & Voter \\
            \hline
            \hline
            \multirow{1}{0.7cm}{Report} & report\_refused\_signature & - & 85$k$ & Voter \\
            \hline
            \hline\\
        \end{tabular}
        \caption{The gas usage of each function in our implementation.}
        \label{tab:gas}
        \label{table:function_gas_usage}
    \end{table}

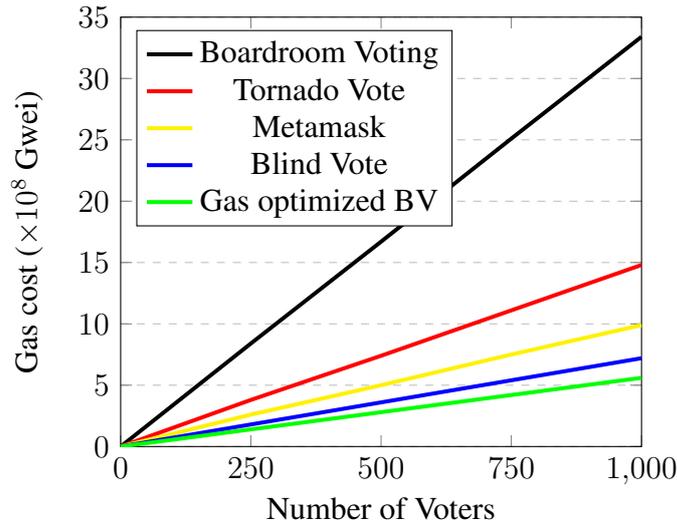
\begin{figure}
    \centering
    \begin{tikzpicture}
        \begin{axis}[
            xlabel={Number of Voters},
            ylabel={Gas cost ($\times 10^8$ Gwei)},
            xmin=0, xmax=1000,
            ymin=0, ymax=35,
            xtick={0,250,500,750,1000},
            ytick={0,5,10,15,20,25,30,35},
            legend pos=north west,
            ymajorgrids=true,
            grid style=dashed,
        ]
        
        \addplot[
            color=black,
            line width=1.5pt,
        ]
        coordinates {
            (0,0)
            (250, 8.4)
            (500, 16.7)
            (750, 25.1)
            (1000, 33.4)
        };

        \addplot[
            color=red,
            line width=1.5pt,
        ]
        coordinates {
            (0,0)
            (250, 3.8)
            (500, 7.4)
            (750, 11.1)
            (1000, 14.8)
        };

        \addplot[
            color=yellow,
            line width=1.5pt,
        ]
        coordinates {
            (0,0)
            (250, 2.6)
            (500, 5.0)
            (750, 7.5)
            (1000, 9.9)
        };

        \addplot[
            color=blue,
            line width=1.5pt,
        ]
        coordinates {
            (0,0)
            (250, 1.8)
            (500, 3.6)
            (750, 5.4)
            (1000, 7.2)
        };

        \addplot[
            color=green,
            line width=1.5pt,
        ]
        coordinates {
            (0,0)
            (250, 1.4)
            (500, 2.8)
            (750, 4.2)
            (1000, 5.6)
        };

        \legend{Boardroom Voting, Tornado Vote, Metamask, Blind Vote, Gas optimized BV}
        \end{axis}
    \end{tikzpicture}
    \caption{Gas comparison of different protocols}
    \label{fig:plot_protocols}
\end{figure}

Figure~\ref{fig:plot_protocols}  compares the gas usage of our approach with 3 previous state-of-the-art blockchain-based voting protocols, namely \cite{Pramulia_Implementation,muth_tornado_vote,mccorry_boardroom}, based on the number $n$ of voters. Our approach significantly outperforms these methods and, assuming that there are $1000$ voters, reduces the gas usage by 43.4\%, 61.9\%, and 83.1\% in comparison to Metamask, Tornado Vote and Boardroom voting, respectively. Among these Tornado Vote is the previous state-of-the-art and the only method that provides the same security guarantees as our approach. Moreover, as Figure~\ref{fig:plot_protocols} shows, the improvement gets more pronounced as the number of voters increases.

\section{Conclusion}

We presented Blind Vote: a secure and gas-efficient approach for secret voting on the blockchain. Blind Vote uses a combination of RSA blind signatures and commitment schemes to attain all the standard desired security properties of a voting protocol, as well as secrecy, i.e.~it is impossible to know which voter cast a particular vote or which vote belongs to a particular voter. We implemented Blind Vote as a free and open-source smart contract and compared its gas usage with previous state-of-the-art blockchain-based secret voting protocols. Blind Vote outperformed these methods significantly in terms of gas usage and obtained improvements of around 40 to 80 percent, hence making blockchain-based secret voting considerably more affordable.

\newpage

\chapter{Blockchain-based Auction}
\label{auction}
This chapter is based on the following publication:

\begin{enumerate}
    \item J. Ballweg, A.K. Goharshady, \textbf{Z. Lin} \textbf{\uline{Fast and Gas-efficient Private Sealed-bid Auctions}} In \textit{44th ACM Symposium on Principles of Distributed Computing} \textbf{(PODC)}, 2025.
\end{enumerate}
\newpage

In this chapter, we introduce an auction protocol that are both gas- and time-efficient. In particular, it costs $O(\log m)$ units of gas per bidder and achieves a runtime of $O(\log m)$. For all the losing bids, this protocol achieves observational determinism as its security guarantees.

\section{Preliminaries and Problem Statement} \label{sec:setting}
In this section, we first outline our setting and assumptions about the blockchain and smart contract environment. These are standard assumptions and can be skipped by readers who are already familiar with this setting. We then define our auction problem and the desired security guarantees. 

\paragraph{Deposits}
Distributed auction protocols often include mechanisms to detect cheating by participants. In a blockchain setting, smart contracts can receive and own money in the form of an underlying cryptocurrency. Thus, we can additionally assume that the participants are required to sign up with the smart contract and pay a deposit to take part in the protocol. This way, any detection of dishonest behavior can immediately be punished by confiscating their deposits.

\paragraph{Identities}
Blockchain environments use asymmetric cryptography and identify users by their public keys. All transactions, including all function calls to smart contracts, have to be signed by the originator. The environment is pseudonymous in the sense that a user can create as many identities as they wish by simply generating more secret/public key pairs. We will exploit this property in our protocol, where users can generate a new identity to send a message to the smart contract without it being connected to their previous identity. In practice, users can have many identities before taking part in the protocol and can use mixing services to fund all of their accounts (public keys) without disclosing their connection to the same person~\cite{mixing,arbabi2023mixing}.

\paragraph{Bidders}
We consider an auction with $n$ bidders numbered from $1$ to $n.$ We use $\pk_i$ to denote the public key (identity) of the $i$-th bidder and $\sk_i$ to denote its corresponding secret key. Naturally, $\pk_i$ is public knowledge whereas $\sk_i$ is only known to $i.$

\paragraph{Auction Problem Statement}
Our goal is to design a sealed-bid auction that can be implemented as a smart contract. Each bidder $i$ should be able to place a bid $b_i$ by an interaction with the auction smart contract. We assume there is a known global maximum bid $m$ and every $b_i$ is between $1$ and $m.$ The contract should then obtain the maximum bid $(\max_i b_i)$ and its bidder $(\argmax_i b_i)$\footnote[2]{If the sequences have several maximal elements, we assume that $\argmax$ returns the set of indices in which the maximum value appears. The protocols are explained as if the auction's winner is unique, but it is trivial to extend them to cases with several winners.}. These values must be obtained in a publicly-verifiable manner, i.e.~anyone with access to the blockchain should be able to verify that the auction has completed successfully and the highest bid/bidder are identified correctly.

\paragraph{Securities Guarantees}
We require our auction to satisfy the following security properties. The first three are classical and can be obtained even by the simplest commit-reveal schemes. Thus, we will mainly focus on the fourth property below:
\begin{enumerate}
    \item \textbf{\emph{Decentralization and Permissionlessness.}} Anyone on the blockchain network can sign up to take part in the auction and no party has permissions to perform operations that are not allowed to any other party.
    \item \textbf{\emph{Trustlessness.}} No party is assumed to be honest. If a party is not following the protocol correctly, this should be identified and punished.
    \item \textbf{\emph{Bid Independence.}} No bidder should be able to change their own bid after learning any information about other bids. This is also called the \emph{sealed-bid} property.
    \item \textbf{\emph{Privacy for Losing Bidders.}} If a bidder $i$ is not the highest bidder, then no information should be leaked about $b_i.$ Note that this property is violated even if $b_i$ or some information about $b_i$ is leaked without it being directly connected to $i.$ For example, if an observer realizes that one of the bids was a particular value, without knowing who the bid belonged to, we still consider this a breach of privacy.
\end{enumerate}

\paragraph{Observational Determinism}
We formalize the fourth property (privacy for losing bidders) above using the notion of observational determinism which is standard in the computer security literature and often used in the context of concurrent programs~\cite{DBLP:conf/csfw/HuismanWS06,DBLP:conf/post/ClarksonFKMRS14,DBLP:conf/csfw/ZdancewicM03,DBLP:journals/jcs/McLean92,DBLP:conf/sp/Roscoe95}. Let $j$ be an observer, i.e.~a user with access to the blockchain network who may or may not be one of the bidders. In one run of the protocol, the observation made by $j$ is the sequence of all transactions and blocks $j$ sees on the blockchain network, which may originate from any participant or miner, together with the timestamps at which they see such transactions/blocks. We denote one such observation with $o.$ Let $B = \langle b_1, b_2, \ldots, b_n \rangle$ be the sequence of bids. We say $o$ is consistent with $B$ from $j$'s point-of-view and write $B \models_j o$ if it is possible that $j$ observes $o$ when the bidders' bids are according to $B.$ Intuitively, if $B \models_j o$ and $B' \models_j o,$ when $j$ observes $o$ they cannot distinguish whether the bids were according to $B$ or $B'.$ We say two bid sequences $B = \langle b_1, b_2, \ldots, b_n\rangle$ and $B' = \langle b'_1, b'_2, \ldots, b'_n\rangle$ are compatible and write $B \compatible B'$ if $\max_i b_i = \max_i b'_i$ and $\argmax_i b_i = \argmax_i b'_i.$

Based on the definitions above, a protocol provides \emph{privacy for losing bidders} if we have:
\begin{itemize}
	\item For every observer $j$ who is not a bidder, if an observation $o$ is consistent with $B,$ then it is consistent with any $B'$ that is compatible with $B.$ Formally,
	$$
	\forall o~~ \forall B~~ \forall B'~~ \left( B \models_j o ~\land~ B \compatible B' \right) \Rightarrow B' \models_j o.
	$$
	\item The same property should hold for every \emph{bidder} $j$ except that the bidder knows their own bid $b_j.$ Thus, if $j$ is a bidder and an observation $o$ is consistent with $B$ from their point-of-view, then it should be consistent with any $B' \compatible B$ as long as $b'_j = b_j.$ Formally,
	$$
	\forall o~~ \forall B=\langle b_1, b_2, \ldots, b_n\rangle~~ \forall B'=\langle b'_1, b'_2, \ldots, b'_n\rangle$$ $$\left(b_j = b'_j ~\land~ B \models_j o ~\land~ B \compatible B' \right) \Rightarrow B' \models_j o.
	$$
	A bidder might take part in the auction with several identities and make several bids. In such cases, we should extend the definition above accordingly to require that the observations match on all bids made by the same person. This also models collusions between bidders.
\end{itemize}
The formal definition above precisely captures our desired privacy concept. Every observer or colluding set of observers would only be able to distinguish between $B$ and $B'$ if they can do so using their own bid(s) and the information about the maximum bid/bidder. Thus, the observation does not leak any information about the losing bids/bidders.

\paragraph{Efficiency Metrics}
To analyze the efficiency of our protocol, we consider the runtime and the maximum gas usage of any bidder. Recall that the runtime is the number of blocks required to execute the protocol. This is not the same as our protocol's computational complexity, given that a single block may contain several transactions/function calls. On the other hand, the gas usage is a closer concept to computational complexity. When a user invokes a function with computational complexity $f,$ it will cost them $\Theta(f)$ in gas. Given that smart contract functions can be called by different users, this cost might be divided among them based on the protocol's requirements. We consider the maximum cost paid by a single user/bidder.

\section{A Dutch Auction with Commitments} \label{sec:dutch}

Our first protocol is a combination of classical commitment-scheme auctions and Dutch auctions. It provides excellent efficiency in terms of gas, requiring only $O(1)$ gas usage for each bidder. Note that this is asymptotically optimal since each bidder must at least sign up with the protocol. However, it requires $\Theta(m)$ time where $m$ is the maximum possible bid. %Thus, it is not practical on its own and will only be used as a subroutine in another auction protocol in Section~\ref{sec:nonbinarytree}. 

\paragraph{Protocol 0. Dutch Auction} In a Dutch auction, an auctioneer starts with a high asking price $m$ and keeps lowering the price until one of the bidders agrees to pay it~\cite{coppinger1980incentives}. This kind of auction originates in Dutch flower markets and can be easily implemented as a smart contact consisting of the following steps:
\begin{enumerate}
	\item[(0)] \textbf{\emph{Parameter Setup.}} The organizer deploys the auction on the blockchain and chooses the deadlines, in terms of block numbers, for each of the following steps. For each step $i,$ the organizer fixes two block numbers $[\tau_i, \tau'_i]$ and the contract accepts function calls of step $i$ only in this period. 
	 The organizer also sets a parameter $d,$ which is the deposit each bidder must pay to join the auction and $m$ which is the maximum allowed bid. This step is the same for all of our future auction protocols and thus we omit it for brevity. Some protocols need additional parameters whose values will also be set in this step.
	 \item[(1)] \textbf{\emph{Registration.}} Anyone on the blockchain network can call a \func{register()} function in this step, paying a deposit $d$. The contract keeps track of the public key $\pk_i$ of every registering bidder.
	 \item[(2)] \textbf{\emph{Countdown.}} This step lasts for exactly $m$ blocks. Throughout this step, any registered bidder can call a function $\func{bid}()$ in the contract. The bidder does not need to provide the value of their bid. In the $x$-th of the $m$ blocks, the contract only accepts bids of value $b_i = m - x + 1.$ Thus, the time of the bid uniquely determines its value\footnote{To avoid issues due to network latency, one can set a longer period of several blocks for each bid value, thus scaling the time by a constant factor.}. The auction concludes as soon as a bid is made. The first bid is automatically the highest bid and its bidder is the auction's winner.
	 \item[(3)] \textbf{\emph{Refund.}} In this step, all bidders can call a \func{refund()} function to receive their deposit $d$ back. Based on the particular use-case, the winner might be excluded.
  \end{enumerate}

Although the simple protocol above provides privacy for losing bidders, whose bids are never revealed, it does not guarantee bid independence. Indeed, when at time $x$ the highest bidder $i$ decides to bid $b_i = m-x+1,$ this is done with the knowledge that no one else's bid is higher than the value $b_i.$ This violates bid independence. For example, a bidder who intends to bid $100$ USD for a batch of Dutch flowers might change their bid to $95$ USD when they realize that no one else has bid $101$ USD or more. To fix this, we combine the classical commit-reveal auction model with the above protocol.

\begin{figure}
    \centering
    \includegraphics[scale=1]{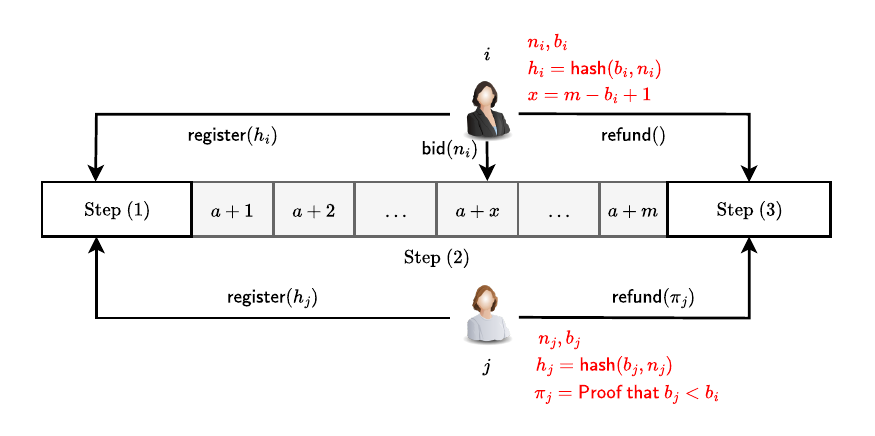}
    \caption{Timeline of function calls by a winning bidder $i$ and a losing bidder $j$ to the auction smart contract in Protocol 1.}
    % In this figure, Step (1) ends at block number $a$ and Step (2) spans blocks $a+1$ to $a+m.$ Red values are private computations made by the bidders on their own machines.
\end{figure}

\paragraph{Protocol 1. Dutch Auction with Commitments} Our new protocol consists of the following steps:
\begin{enumerate}
	\item[(1)] \textbf{\emph{Registration and Commitment.}} Anyone on the blockchain network can register. To do so, they must first commit to their bid $b_i.$ They choose a random nonce $n_i$ and compute $h_i = \func{hash}(b_i, n_i).$ Here, $\func{hash}$ is a cryptographic hash function. They then call $\func{register}(h_i)$ and pay a deposit $d.$ The contract saves the registrant's public key $\pk_i$ and their declared hash $h_i.$
	\item[(2)] \textbf{\emph{Countdown.}} This step lasts for exactly $m$ blocks and is similar to the previous protocol. Throughout this step, any registered bidder $i$ can call a function $\func{bid}(n_i)$ in the contract. The bidder does not need to provide the value of their bid, but only their random nonce $n_i$. In the $x$-th of the $m$ blocks, the contract only accepts bids of value $b_i = m - x + 1.$ When $\func{bid}(n_i)$ is called by bidder $i,$ the contract verifies that $h_i = \func{hash}(b_i, n_i).$ If not, the bidder will be penalized and the function call ignored. As before, the first bid that passes the hash check is automatically the highest bid.
	\item[(3)] \textbf{\emph{Refund.}} In this step, each bidder $j$ can call a function $\func{refund}(\pi_j)$ to receive their deposit $d$ back. The winner might be excluded based on the use-case. Moreover, any non-winning bidder must provide a proof $\pi_j$ showing that their bid $b_j$ was smaller than the winning bid $b_i.$ For this, one does not need to publish the nonce $n_j$ and can use any standard zkSNARK such as Groth16~\cite{DBLP:conf/eurocrypt/Groth16} instead. More specifically, $\pi_j$ is a proof that $j$ knows values $n_j$ and $b_j$ such that $b_j < b_i$ and $\func{hash}(b_j, n_j) = h_j.$ The contract verifies $\pi_j$ and issues the refund only if $\pi_j$ is valid. Otherwise, the bidder's deposit will be burned.
\end{enumerate}

\paragraph{Efficiency} Our protocol above has the same runtime and gas usage as a vanilla Dutch auction. The time is dominated by the countdown which takes $\Theta(m)$ blocks in the worst case. On the other hand, each bidder pays only $O(1)$ in gas, since they have to send a single constant-sized registration transaction in Step (1), followed by a single bid in Step (2) only if they are the winner, and a single refund transaction in Step (3) if they are not the winner, which verifies a constant-sized zkSNARK with constant gas usage.

\paragraph{Security Analysis} Public verifiability is immediate since the contract verifies everything and anyone on the blockchain network can simply run it, too. Decentralization and trustlessness are easy to check. Bid independence is achieved since every bidder commits to their bid in Step (1) and thus cannot change it later. At this point, they have no information about other bids. Privacy for losing bidders is obtained by design since no losing bid is revealed in Steps (2) and (3) and each losing bidder simply provides a zkSNARK certifying they have not won the auction.

\section{Binary Auction Trees} \label{sec:binarytree}

While Protocol 1 of the previous section has all the desired security properties, it takes $\Theta(m)$ blocks to execute. This is prohibitively large for real-world auctions. For example, if we have an auction in which $m=10^6$ and each block takes 13 seconds, as it does on Ethereum, then the countdown step would require almost 150 days. In this section, we provide a protocol that improves the runtime to $\Theta(\log m)$ blocks, albeit at a slightly increased cost of $\Theta(\log m)$ gas for a bidder in the worst case.

\begin{figure}[H]
    \centering
    \includegraphics[scale=1.6]{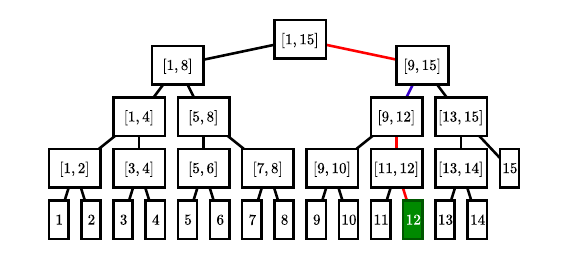}
    \caption{A Binary Auction Tree (BAT) in which $m=15$ and the path taken if the maximum bid is $12.$}
    \label{fig:bat}
\end{figure}

\paragraph{Binary Auction Tree} The idea is to use a binary interval tree of possible bids, which we call a \emph{binary auction tree} (BAT). The root of a BAT corresponds to the interval $[1, m].$ Each vertex $v$ of the tree that is labeled by the interval $[x, y]$ will have two children, the left one corresponding to $\left[x, \lfloor \frac{x+y}{2} \rfloor\right]$ and the right being labeled by the interval $\left[\lfloor\frac{x+y}{2} \rfloor + 1, y\right].$ Each leaf will correspond to a single possible bid value, i.e.~an interval $[x, x]$. For example, Figure~\ref{fig:bat} shows the BAT for $m=15$, the red edges correspond to explicit calls to $\func{right}()$ and the blue edge is an implicit move to the left child.

\paragraph{Intuition} In our second protocol, the smart contract starts at the root of the BAT and keeps traversing it down until it reaches a leaf. At each level of the tree, the bidders must collectively guide the contract towards the highest bid. Since each bidder knows their own bid, they should send a message to the contract if their bid is in the right child of the current node. If the contract does not receive any such message by a particular deadline, it moves to the left child. Otherwise, it moves to the right child. This continues until the contract finds the maximum bid. The actual protocol is a bit more involved as (i)~a smart contract can change states only when one of its functions is called, i.e.~it cannot automatically invoke itself, and (ii)~we need to penalize dishonest bidders.

\paragraph{Protocol 2. Auction using a BAT} Our protocol consists of the following steps:
\begin{enumerate}
	\item[(1)] \textbf{\emph{Registration and Commitment.}} Same as in Protocol 1.
	\item[(2)] \textbf{\emph{Path Finding.}} Suppose the BAT has depth $k,$ i.e.~the distance from the root to the farthest leaf is $k$ edges. This step will take exactly $k$ blocks time. If the previous step ends at block number $a,$ then the current step runs from block $a+1$ to $a+k$\footnote{As in the previous protocols, one can scale this to several blocks for each edge.}. The smart contract implicitly keeps track of its position in the BAT by saving the interval corresponding to the current node, as well the number of steps taken so far. In the time period of block $a+t,$ if the smart contract is in a non-leaf vertex with corresponding interval $[x, y],$ then in the next step it should go to either the left child $\left[ x, \lfloor \frac{x+y}{2} \rfloor \right]$ or the right child $\left[ \lfloor \frac{x+y}{2} \rfloor + 1, y \rfloor \right].$ If a bidder $i$ realizes that their bid $b_i$ belongs to the right child, i.e.~$ \lfloor \frac{x+y}{2} \rfloor + 1 \leq b_i \leq y,$ they should call the function $\func{right}()$ of the smart contract. This call tells the contract to move to the right child. Importantly, since we do not want bidder $i$ to publicly disclose that their bid $b_i$ is in a particular interval, this call is not performed using the identity $\pk_i$ that was used for registering bidder $i$ in Step (1), but instead using several pseudonyms, i.e.~ different identities generated by the same bidder only for this call. The number of such pseudonymous calls is randomly chosen by the caller. Moreover, every call to $\func{right}()$ must have a deposit $d'$ attached to it. The purpose of this deposit will soon become clear. If a $\func{right}()$ transaction is already issued by someone else in the current block, an honest bidder will not make the calls at all. Otherwise, they will make at least $r$ calls to $\func{right}()$. $r$ is a parameter fixed in Step (0). The contract will ignore repeated calls to $\func{right()}$ in the same block and return their deposits\footnote{In practice, the preferred design is to include the interval $[x, y]$ as a parameter and call $\func{right}([x, y])$ to avoid attacks that reuse stale function call transactions.}. 
	
	A further implementation detail in this step is that smart contracts cannot generally invoke their own functions automatically. Thus, moving right is always by a function call from a bidder, but moving left is implicit and is only performed when the next right move is called or the time limit for Step (2) expires and a function in Step (3) is called. For example, a pseudocode for $\func{right}()$ is as follows:
	
\begin{center}
    \fbox{
        \parbox{.7\textwidth}{
            \begin{algorithmic}
                \State step $\gets$ 0
                \State $x \gets 1$
                \State $y \gets m$
                \Procedure{right}{}
                    \State $t \gets \texttt{block.number} - a$
                    \While{step $< t-1 ~\land x \neq y$} \Comment{Implicit left steps}
                        \State $y \gets \lfloor  \frac{x+y}{2} \rfloor$
                        \State step $\gets$ step $+ 1$
                    \EndWhile
                    \If{$x \neq y$} \Comment{Explicit right step}
                        \State $x \gets \lfloor  \frac{x+y}{2} \rfloor + 1$
                        \State step $\gets$ step $+ 1$
                    \EndIf
                \EndProcedure
            \end{algorithmic}
        }
    }
\end{center}

    \paragraph{Example} In Figure~\ref{fig:bat}, if the highest bid is $12,$ then the highest bidder calls $\func{right}()$ in the first block of Step (2) moving to $[9, 15]$. This call is done using a different identity (pseudonym) than the one used for registration. There might be other calls to $\func{right}()$ at this point, too, but the contract will ignore repeated calls. Then, in the second block of Step (2), no one calls $\func{right}().$ Thus, all bidders realize that the contract has implicitly moved to $[9, 12]$ but this change is not yet executed in the contract. In the third block, the highest bidder calls $\func{right}()$ again. At this point, the contract first performs the left move from the last block, going to $[9, 12]$ and then the current right move going to $[11, 12].$ Finally, in the fourth block, the highest bidder calls $\func{right}()$ and the contract ends up in leaf $12.$

	\item[(3)] \textbf{\emph{Revealing the Highest Bid.}} At the end of the previous step, we reach a leaf of the tree that corresponds to a particular bid value $b_i$ belonging to a bidder $i.$ In this step, the winner $i$ must call $\func{bid}(n_i)$ using their original identity $\pk_i$ and provide their nonce $n_i.$ The contract first takes any remaining implicit left steps to find the value $b_i,$ and then checks that $\func{hash}(b_i, n_i) = h_i.$ If no bidder calls $\func{bid}(n_i)$ successfully within the time limit of this step, then the last person to call $\func{right}()$ has been dishonest. In this case, anyone can call a function named $\func{blame()}.$ The contract then confiscates the deposit $d'$ of the last person who moved $\func{right}(),$ goes back to immediately after the second-last call to $\func{right}(),$ or the root if no such call exists, and redoes Step (2) to find a new leaf.
	
	\item[(4)] \textbf{\emph{Refund.}} This step is exactly the same as in Protocol 1. Each bidder $j$ can call a function $\func{refund}(\pi_j)$ to receive their deposit $d$ back. Every non-winning bidder must provide a zkSNARK proof $\pi_j$ showing that their bid $b_j$ was smaller than the winning bid $b_i.$ Additionally, all $d'$ deposits for moving right can be refunded.
\end{enumerate} 

\paragraph{Efficiency} In this protocol, anyone can take part in guiding the contract through the binary auction tree. This is because we want to allow the bidders to use pseudonyms, i.e.~identities other than the ones used in the registration phase, to guide the path. This does not affect our runtime. Note that spurious calls to $\func{right}()$ are disincentivized. If they cause the contract to exceed the actual maximum bid, then they will be punished since each move to the right requires a deposit $d'$. If they do not exceed the maximum bid, they have no effect on the correct execution of the protocol. Thus, in the presence of rational parties, we can analyze the runtime and gas usage with the assumption that we have no such dishonest calls. In this case, the runtime bottleneck is Step~(2) which requires $k = \Theta(\log m)$ blocks. Moreover, each bidder has to pay for gas used in one function call for registration, at most $k$ calls to $\func{right}()$ in Step~(2) and then at most one call in each of Steps~(3) and~(4). Each call uses constant gas. Thus, the worst-case gas usage of a bidder is $\Theta(\log m).$

\paragraph{Further Incentives} In the protocol above, the highest bidder is incentivized to correctly guide the contract to the leaf corresponding to their bid. Otherwise, they will lose their initial deposit $d.$ However, at each step of the path, the protocol requires an honest bidder who wants to go right to submit not just one, but several calls to $\func{right}()$ from different identities. To incentivize this, we can edit the smart contract to remember the first $r$ calls to $\func{right}()$ at each step and later pay a fixed reward to each of them. The reward will be taken from the bidders' initial deposits and its amount, as well as $r,$ are parameters set in Step (0). This way, if several bidders know that we should go right at a particular step, they will compete on sending the information to the contract as soon as possible by creating many calls to $\func{right}()$\footnote{Another implementation detail is that $\func{right}()$ should not allow itself to be called from any other smart contract. Thus, every call to $\func{right}()$ is in a separate transaction.}. 

\paragraph{Security Analysis} Public verifiability, decentralization, trustlessness and bid independence are achieved by arguments similar to the case of Protocol 1. Thus, we focus on analyzing privacy for losing bidders. Let $j$ be a losing bidder. Given that $j$ has signed up in the contract using the identity $\pk_j$ but made calls to $\func{right}()$ by different identities, no one can observe anything about $b_j$ that is connected back to $j.$ This guarantee is sufficient for many real-world use-cases but is weaker than the formalism using observational determinism provided in Section~\ref{sec:setting}. Indeed, this formalism is not satisfied by Protocol 2. As an example, consider the BAT in Figure~\ref{fig:bat} and suppose the highest bidder $i$ is bidding $b_i = 12.$ Suppose $j$ is the second-highest bidder and $b_j = 10.$ In the first step, $i$ plans to call $\func{right}()$ but observes that someone else calls $\func{right}()$ first. This tells $i$ that there is at least one other bid in the range $[9, 15].$ Although the identity of $j$ is kept secret, the information that the second highest bid is also in $[9, 15]$ is enough to violate our strict privacy requirement. Of course, when $i$ wins the auction, they will know that the second highest bid is in $[9, 12].$

\section{Fake Bids and Observational Determinism} \label{sec:fakebids}

While our Protocol 2 does not satisfy observational determinism as defined in Section~\ref{sec:setting}, it comes quite close to it. There is no link between the right moves on the tree and the original identities of the bidders. Additionally, any leaked information is only about the second-highest bid. Intuitively, if the highest bidder plans to move right at some point but observes that someone else made the move first, they will know that the second-highest bid is in the right subtree, too. However, such an observation can be made irrespective of the other bids and thus does not leak any information about them. From the point-of-view of anyone other than the highest bidder, observational determinism is already satisfied. If an observer $j$ who is not the highest bidder observes several calls to $\func{right}()$ at a particular block, they cannot know if the calls originated from the same person who is using several pseudonyms or a number of different people. This was the reason behind issuing each $\func{right}()$ calls many times. Thus, they only gain information about the highest bid, which is not a violation of our privacy requirements.

\paragraph{Protocol 3. Auction using a BAT and Fake Bids} To provide privacy for the second-highest bidder and ensure the desired observational determinism from the point-of-view of the highest bidder, we simply allow $f$ of the bidders to each make an additional fake bid. $f$ is a parameter set in Step~(0). Specifically, our Protocol 3 has the following steps:
\begin{enumerate}
    \item [(1)] \textbf{\emph{Registration and Commitment.}} Same as in Protocols 1 and 2.
    \item [(2.1)] \textbf{\textit{Selecting Fake Bidders.}} $f$ of the $n$ bidders are selected randomly to be \emph{fake bidders}. To choose the fake bidders, the smart contract relies on the output of a blockchain-based random number generator such as Randao~\cite{randao}. Random number generation is a well-studied topic in blockchain and there are many efficient, tamper-proof and secure solutions available~\cite{DBLP:conf/icbc2/ChatterjeeGP19,DBLP:conf/sp/SchindlerJSW20,fatemi2025fortuna, abidha2024gas, barakbayeva2024srng, fatemi2023secure, cai2023trustless, chatterjee2019probabilistic}. Specifically, if the output of the RNG service in the previous block is $\rho$ and $\func{rand}$ is a pseudo-random number generator, then $\rho$ is chosen as the seed and $\func{rand}$ is called $f$ times to choose the $f$ fake bidders. In implementation, $\func{rand}$ can be instantiated to any cryptographic hash function.
    \item [(2.2)]\textbf{\textit{Computing Fake Bids.}} If a bidder $i$ is chosen as a fake bidder, they first generate a fake bid 
        \begin{equation} \label{eq:fakebid} b'_i = \left(\func{rand}(\rho, n_i) \mod m\right) + 1.\end{equation}
     Note that the fake bid is uniformly distributed and depends on the RNG output $\rho$ and the bidder's nonce $n_i.$ They then take part in the protocol with both original and fake bids, i.e.~$b_i$ and $b'_i.$
    \item [(2.3)] \textbf{\textit{Path Finding.}} This step is exactly the same as Step (2) of Protocol 2, except that the fake bidders call $\func{right}()$ whenever either of their two bids is in the right subtree.
    \item [(3)] \textbf{\emph{Revealing the Highest Bid.}} At the end of the path-6finding step, we reach a leaf of the tree. If the leaf corresponds to a real bid $b_i,$ then the bidder $i$ must call $\func{bid}(n_i).$ The contract verifies this as in the previous protocol. Otherwise, if no such bid is declared, the leaf corresponds to a fake bid $b'_i.$ In this case, the fake bidder $i$ must call $\func{fakebid}(\pi'_i)$ and provide a zkSNARK $\pi'_i$ proving that their fake bid $b'_i$ is indeed the leaf we have ended up in. Given that $\rho$ is public knowledge, $\pi'_i$ will simply be a proof that $i$ knows a value $n_i$ which satisfies equation~\eqref{eq:fakebid}.
    \item [(4)] \textbf{\emph{Verification of Fake Bids.}} Every fake bidder $j$ must call $\func{fakebid}(b'_j, \pi'_j),$ providing their fake bid $b'_j$ and a zkSNARK proof that it was computed according to equation~\eqref{eq:fakebid}. If $b'_j$ is larger than the leaf reached in Step (3) or $\pi'_j$ is invalid or not provided, the contract confiscates the deposit of $j$ and disallows them from continuing to participate in the auction.
    \item [(5)]\textbf{\emph{Refund or Reset.}} At this step, anyone who has paid a deposit to call $\func{right}()$ can take the deposit back. If the leaf identified in Step (3) corresponds to a real bid, then the refund step is triggered, which works exactly as in Protocol 2, i.e.~every non-winning bidder $j$ can call $\func{refund}(\pi_j),$ providing a zkSNARK $\pi_j$ proving that their bid was smaller than the maximum bid $b_i$ identified in Step (3) and receiving their deposit back. However, if the leaf identified in Step (3) corresponds to a fake bid, i.e.~if the maximum bid $b'_i$ is fake, the smart contract sets $m \gets b'_i - 1,$ goes back to Step (2.1) and performs a new walk from the root of the binary auction tree to a leaf.
\end{enumerate}

\paragraph{Efficiency} The runtime and gas usage of Protocol 3 are similar to Protocol 2, except that the root-to-leaf walks on the BAT may be repeated several times. Specifically, suppose $m$ is the maximum allowed bid and $b_i$ is the largest bid. $m$ decreases in each round, but we ignore this for ease of analysis. The probability that no fake bids surpass $b_i$ is at least $\left(b_i/m\right)^f.$ Thus, the expected number of times the protocol has to traverse a root-to-leaf path is less than $\left({m}/{b_i}\right)^f.$ Therefore, the expected runtime of the protocol and the expected gas usage per bidder are both in $O\left(\left(m/b_i\right)^f \cdot \log m\right).$ If both $f$ and $m/b_i$ are small constants, i.e.~only a few fake bids are allowed and the maximum allowed bid $m$ is chosen reasonably so that it does not exceed the real maximum bid by more than a constant factor, then the asymptotic performance matches that of Protocol 2, i.e.~$O(\log m)$ runtime and $O(\log m)$ gas usage per bidder.

\paragraph{Security Analysis} We only need to prove observational determinism since the other desired properties are inherited from Protocol 2. We assume that the constant $f > 1$ is chosen in a way that no one person may control all $f$ fake bidders. In practice, since taking part in the protocol is costly due to the deposit and gas payments, even a small $f$ suffices. Consider an observer $j$ who makes an observation $o$ during one root-to-leaf round of Protocol 3. $o$ may contain several pseudonymous calls to $\func{right}().$ Since these calls are pseudonymous, $j$ is unable to connect any of them to another bidder $i.$ Moreover, $j$ cannot obtain information about any of the bid values, except their own value $b_j$ if they are a bidder. This is because $j$ cannot distinguish the calls to $\func{right}()$ that are made due to a real bid from those that are due to a fake bid. More specifically, if the round ends at a leaf that is then revealed in Step (3) as a real bid $b_i$ with $i \neq j,$ then it is possible that all the calls to $\func{right}()$ in the current path were invoked by $i$'s pseudonyms. Thus, from $j$'s perspective, the observation is consistent with any bid sequence in which the maximum element is $b_i.$ Alternatively, if $j$ is the maximum bidder and in Step (3) $b_j$ is revealed as the maximum bid, then the observation $o$ is still consistent with any bid sequence in which the maximum element is $b_j.$ This is because there might be a different bidder $i$ whose fake bid is $b'_i = b_j.$ This fake bidder would call $\func{right}()$ according to the same path as $j$ but will not reveal a real bid in Step (3). Finally, if the bid revealed in Step~(3) is fake, the exact same argument establishes observational determinism, i.e.~the observation is consistent with any sequence of bids whose maximum is less than the revealed fake bid of Step~(3).

\section{Conclusion}

In this work, we presented a novel blockchain-based protocol for first-price sealed-bid auctions that guarantees privacy for losing bidders, i.e.~that their bids are not leaked as formalized by the concept of observational determinism. Our protocol can be implemented as a smart contract on any programmable blockchain and is efficient in terms of both time and gas. It concludes within $O(\log m)$ blocks, where $m$ is the maximum allowed bid, and each bidder pays an expected gas cost of $O(\log m).$ A limitation of our approach is that observational determinism models non-determinism in a system but does not consider probabilistic behavior or inference. Extending the auction protocol with a stronger probabilistic security guarantee, such as those provided by zero-knowledge protocols, is an interesting direction of future work.
\newpage

\addcontentsline{toc}{chapter}{Bibliography}
\bibliographystyle{IEEEtran}
\bibliography{references}
\newpage

%%%%%%%%%%%%%%%%%%%%%%%%%%%%%%%%%%%%%%%%%%%%%%%%%%%%%%%%%%%%%%%%%%%%%%%%%
%                                                                       %
%     10) APPENDIX (If Any)                                              %
%                                                                       %
% \appendix command marks the beginning of the APPENDIX part of the     %
% Thesis. The usual \chapter command is used for the different chapters %
% of the Appendix.                                                      %
%                                                                       %
%%%%%%%%%%%%%%%%%%%%%%%%%%%%%%%%%%%%%%%%%%%%%%%%%%%%%%%%%%%%%%%%%%%%%%%%%
% \appendix
% \chapter{Tips on HKUST thesis preparation}
% \label{chp_tips}
% \input{A_tips.tex}

%%%%%%%%%%%%%%%%%%%%%%%%%%%%%%%%%%%%%%%%%%%%%%%%%%%%%%%%%%%%%%%%%%%%%%%%%
%                                                                       %
%     11) BIOGRAPHY (optional)                                          %
%                                                                       %
% \biography and \endbiography are used to define the optional          %
% Biography of the author of the Thesis.                                %
%                                                                       %
%%%%%%%%%%%%%%%%%%%%%%%%%%%%%%%%%%%%%%%%%%%%%%%%%%%%%%%%%%%%%%%%%%%%%%%%%

\end{document}